%% file: jstsp_v0.tex
\documentclass[journal]{IEEEtran}

\usepackage{amssymb}
\usepackage[cmex10]{amsmath}
\usepackage{stfloats}
\usepackage{graphicx}
\usepackage{subfigure}
\usepackage{tabularx}
\usepackage{booktabs}
\usepackage{threeparttable}
\usepackage{epsfig,epsf,color,balance,cite}
\usepackage{verbatim}
\usepackage{url}
\usepackage{bm}

\usepackage{algorithm}
\usepackage{algorithmic}
\usepackage{multirow}

\usepackage{longtable}

\usepackage{graphicx}
\usepackage{multirow}
\usepackage{multicol}

\usepackage{supertabular}
\usepackage{longtable}
\begin{document}

\title{Edge Learning for B5G Networks with Distributed Signal Processing: Semantic Communication, Edge Computing, and Wireless Sensing}


\author{Wei Xu,~\IEEEmembership{Senior Member,~IEEE},
Zhaohui Yang,~\IEEEmembership{Member,~IEEE},
Derrick Wing Kwan Ng,~\IEEEmembership{Fellow,~IEEE}, \\
Marco Levorato,~\IEEEmembership{Senior Member,~IEEE},
Yonina C. Eldar,~\IEEEmembership{Fellow,~IEEE}, and
M\'erouane Debbah,~\IEEEmembership{Fellow,~IEEE}\\

\thanks{W. Xu is with the National Mobile Communications Research Lab, and Frontiers Science Center for Mobile Information Communication and Security, Southeast University, Nanjing 210096, China, and also with Purple Mountain Laboratories, Nanjing 211111, China (email: wxu@seu.edu.cn).}
\thanks{Z. Yang is with the Department of Electronic and Electrical Engineering, University College London, WC1E 6BT London, UK (email: zhaohui.yang@ucl.ac.uk).}
\thanks{D. W. K. Ng is with the School of Electrical Engineering and Telecommunications, The University of New South Wales, Australia (email:w.k.ng@unsw.edu.au).}
\thanks{M. Levorato is with the Department of Computer Science, University of California, Irvine, CA 92697, USA (email: levorato@uci.edu).}
\thanks{Y. C. Eldar is with the Faculty of Math and CS, Weizmann Institute of
Science, Rehovot 7610001, Israel (email: yonina.eldar@weizmann.ac.il).}
\thanks{M. Debbah is with the Technology Innovation Institute and also with the Mohamed Bin Zayed University of Artificial Intelligence, 9639 Masdar City, Abu Dhabi, United Arab Emirates (email: merouane.debbah@tii.ae).}
}
%

\maketitle

\input{./sections/abstract}

\IEEEpeerreviewmaketitle

\input{./sections/main}


\bibliographystyle{IEEEtran}
\bibliography{MMM}

\end{document}

%% file: sections/abstract.tex
\begin{abstract}
    To process and transfer large amounts of data in emerging wireless services, it has become increasingly appealing to exploit distributed data communication and learning. Specifically, edge learning (EL) enables local model training on geographically disperse edge nodes and minimizes the need for frequent data exchange. However, the current design of separating EL deployment and communication optimization does not yet reap the promised benefits of distributed signal processing, and sometimes suffers from excessive signalling overhead, long processing delay, and unstable learning convergence. In this paper, we provide an overview on practical distributed EL techniques and their interplay with advanced communication optimization designs. In particular, typical performance metrics for dual-functional learning and communication networks are discussed. Also, recent achievements of enabling techniques for the dual-functional design are surveyed with exemplifications from the mutual perspectives of ``communications for learning'' and ``learning for communications.'' The application of EL techniques within a variety of future communication systems are also envisioned for beyond 5G (B5G) wireless networks. For the application in goal-oriented semantic communication, we present a first mathematical model of the goal-oriented source entropy as an optimization problem. In addition, from the viewpoint of information theory, we identify fundamental open problems of characterizing rate regions for communication networks supporting distributed learning-and-computing tasks. We also present technical challenges as well as emerging application opportunities in this field, with the aim of inspiring future research and promoting widespread developments of EL in B5G.  
\end{abstract}


%% file: sections/main.tex
\input{./sections/section1-introduction}

\input{./sections/section2-A}
\input{./sections/section2-B}
\input{./sections/section2-C}

\input{./sections/section3-A}
\input{./sections/section3-B}
\input{./sections/section3-C}

\input{./sections/section4-A}

\input{./sections/section4-B}

\input{./sections/section4-C}
\input{./sections/section4-D}

\input{./sections/section4-E}

\input{./sections/section5}

\input{./sections/section6-conclusions}

%% file: sections/section1-introduction.tex
\section{Introduction}


\subsection{Motivation of Edge Learning}

Owing to the massive amount of data traffic
for the role-out of 
the Internet-of-Everything (IoE), machine learning (ML) is envisioned to be an important technology to facilitate the evolution of beyond 5G (B5G) networks \cite{saad2019vision}. Traditional ML methods needs to centrally train data on a specific data center \cite{liu2021learning,liu2021deep,gao2019deep,gao2020unsupervised}. However, owing to the privacy concern and shortened wireless communication resource to support extensive data transfer, all edge devices cannot transmit the data that they have collected to a data center to execute centralized ML methods for data processing. This has triggered the fast-growing research field, namely edge learning (EL), which can deeply integrate two main directions: wireless communications and ML. Advances in EL are widely expected to provide a platform to implement the edge artificial intelligence (AI) in B5G networks \cite{chen2020joint,Bennis2020Communication,letaief2021edge,wang2022asynchronous}.

\subsection{Edge Learning in B5G Networks}

The EL framework allows distributed ML over numerous edge devices that are controlled through multiple wireless servers to collaboratively train massive AI models utilizing the local data and distributed processors, e.g., central processing units (CPUs) and graphic processing units (GPUs) \cite{konevcny2016federated,zhu2019broadband}.
Compared with distributed ML, EL refers to that multiple edge devices cooperatively train the ML model and this process is implemented over edge networks.
The process of EL necessitates the download and upload of large-dimension ML parameters as well as their frequent updates among multiple edge devices. These new paradigms are expected to generate enormous data traffic, which can 
increase burden to the already congested communication networks \cite{yang2021federated}. This challenging issue cannot be addressed by using current wireless techniques aiming at  capacity maximization, as they are decoupled from ML. Realizing the goal of EL with high communication efficiency requires advanced techniques of new distributed signal processing and wireless techniques that seamlessly 
integrate communications and learning approaches.

The deployment of EL in B5G networks leads to dual-functional performance metrics  for both learning and communication. 
On the one hand, the EL framework requires frequent parameter exchanges among edge devices or between  edge devices and a central aggregator through capacity-limited wireless links. 
Thus, wireless communication resource allocation, such as beamforming design, power control, user scheduling, and resource block allocation, can be optimized to improve the dual-functional performance metrics to facilitate learning  \cite{chen2020joint,Bennis2020Communication,chen2021communication}.
On the other hand, the spectral and energy  efficiency optimization of B5G networks often results in less tractable nonconvex resource allocation problems due to interference \cite{liu2017nonorthogonal,liu2021learning}. Traditional signal processing algorithms relying on a local search can only guarantee a sub-optimal solution and centralized learning techniques usually lead to high communication signaling overhead and long delay. Thus, EL, in the form of distributed reinforcement learning (DRL) represents an elegant and efficient mechanism to enable distributed optimization procedure to approach the optimal solution of wireless resource allocation problems requiring only limited overhead under stringent  delay constraints.

\subsection{Focus and Structure}
There are some recent surveys about EL techniques.
For instance, in \cite{verbraeken2020survey} the opportunities and advantages of distributed and centralized ML algorithms were discussed from the viewpoint of computer science. 
Moreover, the authors in \cite{yang2019federated,imteaj2021survey,wahab2021federated} covered the technical issues and recent progress of a specific EL framework of federated learning (FL).
Additionally, possible architectures of EL over wireless communication networks were summarized in \cite{chen2021distributed,Latief2021EdgeAI}. 
Compared with these above works \cite{verbraeken2020survey,yang2019federated,imteaj2021survey,wahab2021federated,chen2021distributed,Latief2021EdgeAI}, the main focus of this paper is to provide a comprehensive overview of state-of-the-art signal processing techniques for EL over B5G networks. 


We aim to gather recent contributions that address the key challenges of applying EL
techniques to understanding and designing upcoming B5G networks from the viewpoint of joint learning and communication. In particular, our objectives are two-fold: 1) to provide
the key open problems in B5G raised in the applications of EL methods, and 2) to pinpoint
main EL techniques that can be adopted for developing B5G.

\begin{figure*}[t]
\centering
\includegraphics[width=7in]{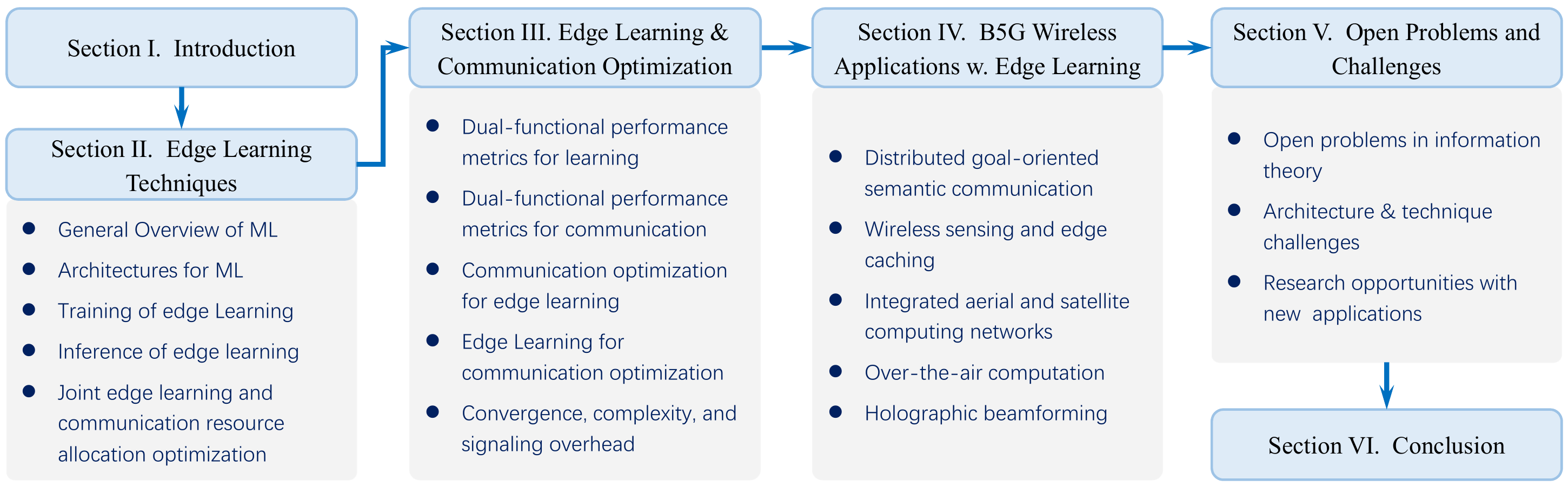}
\caption{Structure of this paper: An overview of the main results and topics.} \label{paperstructure}
\end{figure*}

In the rest of this paper, we first provide an overview of EL techniques from the viewpoint of joint learning and communication in Section II. Then, in Section III, interplay between EL and wireless communication systems is introduced in detail, including dual-functional performance metrics and optimization frameworks. Emerging applications of EL in B5G networks are further discussed in Section IV.  Finally in Section V, open problems and challenges are pointed out before the concluding remarks in Section VI.
The structure of this paper is summarized in  Fig.~\ref{paperstructure}.
Meanwhile, in Table~\ref{KeyAcronym}, we list the key acronyms about ML used in this paper. 
\begin{table}[]
\centering
\caption{Key Acronyms in Learning}
\label{KeyAcronym}
\setlength{\tabcolsep}{5.2mm}
\begin{tabular}{|c|l|}
\hline
\textbf{Acronym}   & \textbf{Description}                        \\ \hline
AI           & Artificial Intelligence            \\ \hline
AirComp      & Over-the-air Computation           \\ \hline
Air-FL       & Over-the-air Federated Learning    \\ \hline

CI            & Centralized Inference       \\ \hline
CNN          & Convolution Neural Network     \\ \hline


DI           & Distributed Inference         \\ \hline
DL           & Deep Learning                 \\ \hline
DNN          & Deep Neural Network           \\ \hline
DP           & Differential Privacy          \\ \hline
DQL          & Deep Q-learning               \\ \hline
DRL          & Deep Reinforcement Learning   \\ \hline
EL           & Edge Learning                \\ \hline
           
FDRL         & Federated Deep Reinforcement Learning          \\ \hline
FL           & Federated Learning            \\ \hline
FTL          & Federated Transfer Learning   \\ \hline
GNN          & Graph Neural Network   \\ \hline
HFL          & Horizontal Federated Learning \\ \hline




MARL         & Multi-Agent Reinforcement Learning            \\ \hline
MADRL        & Multi-Agent Deep Reinforcement Learning       \\ \hline
MAFRL        & Multi-Agent Federated Reinforcement Learning  \\ \hline
MDP          & Markov Decision Process       \\ \hline
ML           & Machine Learning              \\ \hline




RL           & Reinforcement Learning        \\ \hline
RNN          & Recurrent Neural Network      \\ \hline

SG          & Stochastic Game             \\ \hline
SGD          & Stochastic Gradient Descent   \\ \hline
SL           & Split Learning               \\ \hline
VFL          & Vertical Federated Learning   \\ \hline

\end{tabular}
\end{table}

%% file: sections/section2-A.tex
\section{Edge Learning Techniques}

The main task of EL is to deploy ML algorithms at network edges such that highly-distributed real-time data generated by edge devices can be used for fast and cost-effective AI training~\cite{zhu2020toward}. In this section, we provide a general overview of major ML techniques and then introduce ML architectures from the perspective of network topology, with special focus on distributed EL. Typical EL training methods including FL, split learning (SL), and multi-agent reinforcement learning~(MARL). The interplay of those EL methods with wireless communications is also briefly discussed to highlight the necessity of communication theory for EL  in turns of its fundamental privacy concerns,  security guarantees, and performance improvement.
\subsection{General Overview of Machine Learning}

In definition, ML methods refer to a set of algorithms that make decisions, inferences, or predictions based on the observed data~\cite{murphy2012machine}. An ML problem can be generally divided into two phases: a training phase and an inference phase. The training phase is used for training particular ML models by utilizing a large amount of data and some specific ML algorithms. The output of the training phase is a trained model. As for the inference phase, the trained model is deployed to support real-world applications, taking new data as input and yielding corresponding inference results. 
Training an ML model requires some form of feedback to guide the learning process. According to the types of feedback, ML algorithms are usually divided into the following paradigms~\cite{verbraeken2020survey}. 
\begin{itemize}
	\item {\bf Supervised learning}. The training data set for this paradigm contains both inputs and labelled outputs. Supervised learning algorithms learn the underlying mapping between the inputs to the outputs. The outputs are also known as (a.k.a.) labels which provide supervised feedback. 
	\item {\bf Unsupervised learning}. The training set for unsupervised learning contains only inputs, without labelled outputs. Unsupervised learning algorithms aim to learn functions that describe intrinsic structural characteristics of the data. Unsupervised learning algorithms have been widely used, for instance, for dimensionality reduction and data clustering~\cite{celebi2016unsupervised}. 
	\item {\bf Semi-supervised learning}. Under the assumption of label sharing among similar data, semi-supervised learning assigns known labels to unlabeled data, e.g., via clustering. In order to minimize the requirement of manual labeling, semi-supervised learning adopts a small labeled dataset and a large amount of unlabeled data, which is more economical than the fully supervised learning methods, while often achieving comparable performance.
	\item {\bf Reinforcement learning}. Different from the above paradigms of learning, the feedback of reinforcement learning~(RL) takes the form of a reward function, which is designed to evaluate the states of a given environment. RL algorithm learns by using agents taking actions based on the observations from the environment. 
\end{itemize}

Along with these mature learning paradigms, a successful ML algorithm also requires massive data and computing power for effective learning. 
Historically, conventional ML algorithms were limited by computing power and the amount of data. As such, shallow structures were used in ML to limit the model complexity. Such shallow structures mostly relied on effective features selected or extracted by human experts, which restrict their learning power in challenging problems, e.g., computer vision and natural language processing~\cite{goodfellow2016deep}. 

Benefiting from developments of high-performance computing hardware and exponentially growing volume of data, it has become now possible to train and deploy more complicated deep structures, e.g., deep neural network~(DNN). One representative branch of ML, namely deep learning~(DL)~\cite{lecun2015deep}, leverages data-driven feature extraction with deep structures of neural networks to achieve performance that approaches, or even surpasses, human skill on tasks such as image classifications~\cite{he2015delving}, machine translations~\cite{wu2016google}, and  gaming~\cite{silver2016mastering}. In addition, DL with powerful feature extraction ability has recently been applied to wireless communications and IoE applications, e.g., physical layer authentication~\cite{merchant2018deep, peng2019deep, xie2021generalizable}, channel state information~(CSI) compression~\cite{sun2020ancinet, yin2022deep, lu2019mimo}, signal detection~\cite{shlezinger2019viterbinet, khobahi2021lord, shlezinger2020deepsic, farsad2021data}, and transceiver optimization~\cite{hu2021iterative, zhang2022data}. However, the increasing complexity of DL applications poses new challenges toward practical system deployments due to computing and storage limitations, especially in processing centralized networks with massive nodes exogenous data, and thus requiring frequent communications.

\subsection{Architectures for Machine Learning}
The complexity of DL has raised an unprecedented growth in demanding computing power and storage resources. There are two main approaches to supply resources to an ML system: 1) scale-up, i.e., allocating more computing and storage resources to a single commodity server, and 2) scale-out, i.e., involving additional compute-capable nodes in the system. Since the growth of data processing requirement of DL training has far exceeded the development of computing power, scale-out has  become a more economical option, motivating ML systems to evolve from a centralized implementation to a distributed realization.
In the following, we introduce the architectures of both centralized and distributed ML systems from the perspectives of both system topology and parallelization.

\begin{figure*}[t]
\centering
\includegraphics[width=7in]{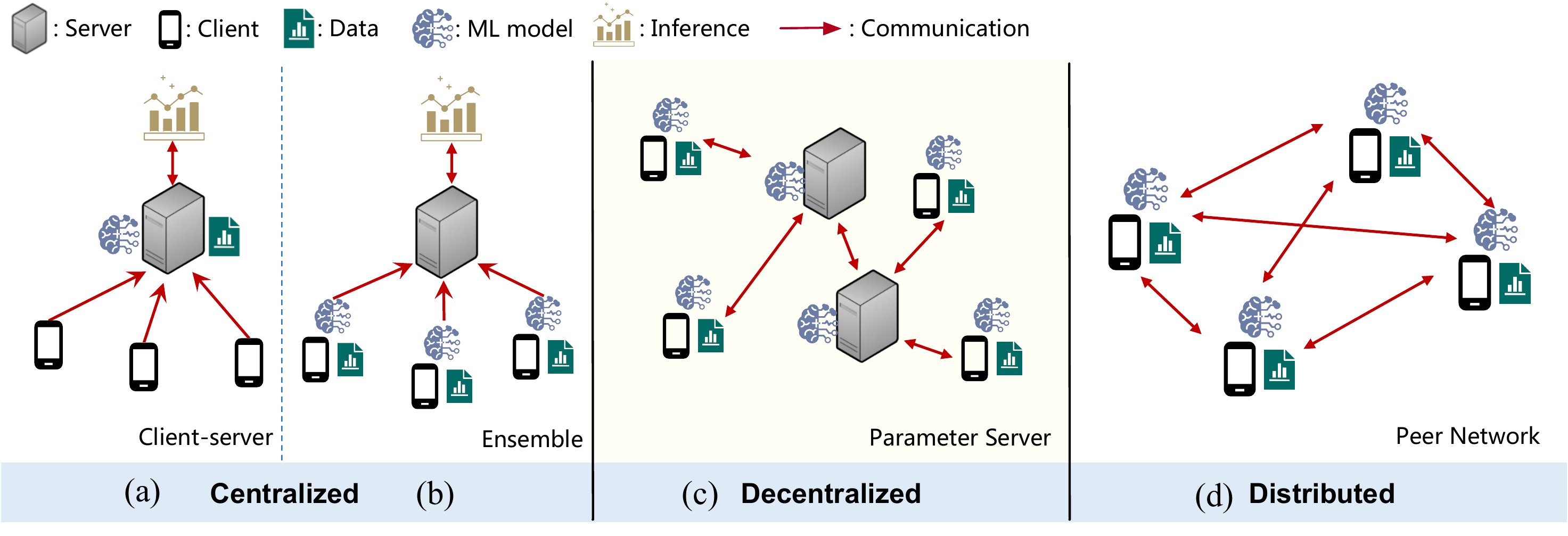}
\caption{Three typologies of distributed ML architectures.} \label{fig2}
\end{figure*}

\paragraph{Topology} Begin with the topology, i.e., the organization of the compute nodes within a learning system. Fig.~\ref{fig2}(a) depicts a conventional client-server topology with a single central server as the only compute-capable node. Data collected from clients are first uploaded to the central server. The central server stores and processes the data. Then, the server returns inference results to the clients. Due to the need for centralized processing of the data, long latency and large transmission costs are incurred when the communication links between the clients and the server have low capacity, or when the clients and server are topologically distant in the network. Furthermore, constraints in computing power and storage resource of the central server introduces challenges when centralized learning are used to support the training of sophisticated models based on extensive datasets. To address these challenges, distributed ML systems have been proposed. We summarize three types of topologies for developing distributed ML according to the degrees of distribution as characterized in \cite{verbraeken2020survey}. 
\begin{itemize}
	\item {\bf Centralized learning architecture}, a star-like topology, refers to a distributed ML system with a strict hierarchical structure and a central aggregation server. Besides the conventional client-server architecture in Fig.~\ref{fig2}(a), another representative learning architecture is \emph{ensemble learning}~\cite{dong2020a}, which is shown in Fig.~\ref{fig2}(b). In ensemble learning, the model training of each node adopts its local data, and the results from the local models are then aggregated on a centralized server using ensemble methods to calculate a global result. This topology, illustrated in Fig.~\ref{fig2}(a)-(b), is easy to deploy and maintain, and is especially suitable for settings where data is scattered across different regions and data interactions are  costly. However, due to the use of local data, the performance of the model on a single node is often unsatisfactory and global calculation are critical for inference, which results in large latency.
	\item {\bf Decentralized learning architecture} includes multiple ``central'' servers and can shape in multiple topologies, e.g., a tree, a ring, and a mixture of both, allowing information aggregation at different levels to synchronize model parameters, as shown in Fig.~\ref{fig2}(c). Decentralized learning architectures, e.g., \emph{AllReduce}~\cite{agarwal2014reliable} and \emph{Parameter Server}~\cite{li2013parameter,li2014scaling,wei2015managed}, have been widely used for large-scale training of DL algorithms. In \emph{AllReduce}, the topology of compute nodes forms a tree structure. Each children node in the tree computes local gradients, aggregates them, and transmits the aggregated gradients to its parent node to complete the gradient calculation. \emph{Parameter Server}, is the prototype of FL~\cite{konevcny2016federated} and implements local computation and global parameter sharing through a set of worker nodes and a set of master nodes. The advantage of Parameter Server is that global data knowledge sharing can be achieved without transferring raw data from local storage. However, the requirement for global model synchronization leads to distributed acceleration bottlenecks. For example, when the computing power of the worker nodes is unbalanced or the worker nodes are heterogeneous, the time consumption of the global model synchronization depends on the slowest compute node, resulting in the computation idleness of the faster compute nodes.
	\item {\bf Distributed learning architecture}, a mesh topology, generally composed of multiple independent compute nodes, with no role differences in the topology and using point-to-point communications (see Fig.~\ref{fig2}(d)). All the nodes own a copy of the model and altogether build a complete solution. This architecture has obvious advantages over the centralized counterpart in terms of scalability and elimination of single points of failure~(SPoF). The challenge is that it results in an extremely high data volume to be transferred for model synchronization.
\end{itemize}

The purpose of distributed ML architectures is to offload computing requirements to multiple compute nodes while considering the communication overhead of model synchronization and data transmission, thereby reducing service latency and computing idle. However, when distributed ML is deployed on wireless devices, limited wireless resources causes additional challenges to learning, such as higher data aggregation error and delay.

\paragraph{Parallelizations}  Another perspective for the design of distributed learning systems is parallelization. In essence, there are two distinct ways, i.e., data parallelism and model parallelism, to split an ML problem across compute nodes~\cite{goodfellow2016deep}.
\begin{itemize}
	\item {\bf Data parallelism.} Based on the assumption of independent and identically distributed~(i.i.d.) data, data parallelism uniformly distributes data to all compute nodes. Additionally, all the nodes share the same algorithmic model through centralization or replication to process different subsets of the data. This  design naturally guarantees that the computing process of the model is consistent with its centralized counterpart. 
	\item {\bf Model parallelism.} The ML model is split into multiple submodels, each of which is deployed on a compute node, such that each node has an accurate copy of the complete data. However, this approach is unsuitable for ML algorithms with non-separable parameters. 
\end{itemize}

Note that the two types of parallelization are not mutually exclusive and they can be used simultaneously in a distributed ML system for flexible deployment.

%% file: sections/section2-B.tex
\subsection{Training of Edge Learning}
\subsubsection{Federated Learning}
\input{./tables/table3}

\begin{figure}[!t]{}
\centering
\includegraphics[width=3.5in]{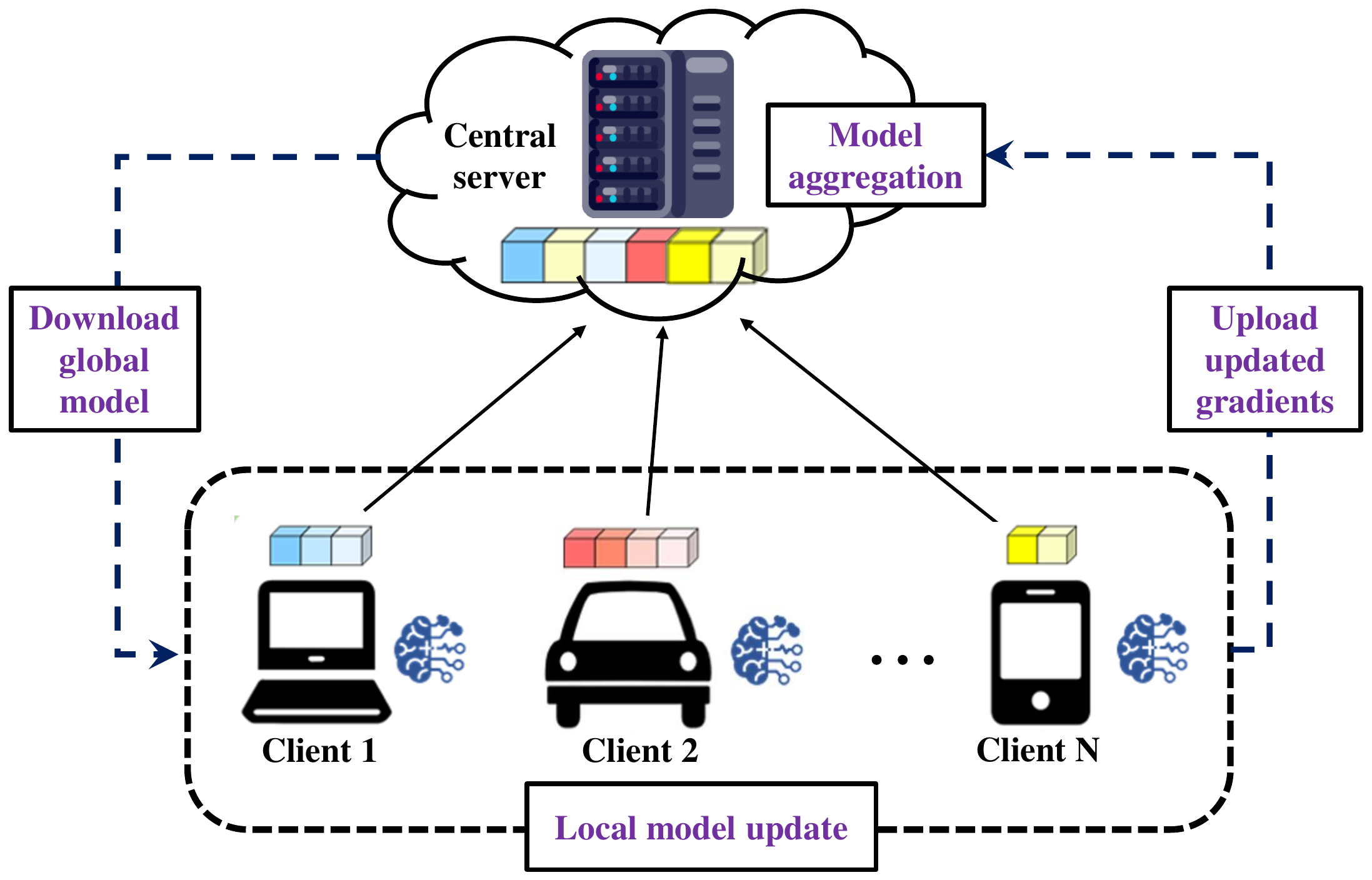}
\caption{Architecture for FL.}
\label{FL}
\end{figure}

FL is a distributed collaborative AI method first proposed by Google in 2016 \cite{konevcny2015federated,mcmahan2017communication}. The main idea of FL is to establish a global ML model based on distributed datasets, where the devices send their local models to the central server without sharing any raw training data.
In general, a FL system consists of two main entities: a central server and a set of clients, denoted by $\mathcal{N}$ \cite{wahab2021federated}. Each client $n\in {\mathcal N}$ owns a local dataset ${\mathcal D}_n=\{\mathbf{X}_n,\mathbf{Y}_n\}$, where $\mathbf{X}_n$ is the feature space vector of client $n$ and $\mathbf{Y}_n$ is the associated label matrix. For each episode, a subset of clients ${\mathcal C}\subseteq {\mathcal N}$ is chosen to participate in the federated training process. Each client $c\in {\mathcal C}$ utilizes its local dataset to independently train and update local gradients. The trained local gradients are then uploaded to the central server for updating the global model. The central server synchronizes the global model, i.e., the weight matrix $\bf{W}$ of a neural network, to all participating clients in ${\mathcal C}$. In the training process of FL, the federated optimization objective is formulated as
\begin{equation}\label{Federated optimization objective}
\underset{{\bf {W}} }{\rm minimize}\quad F({\bf{W}})=\sum\limits_{c=1}^{C}\frac{s_c}{s}f_c({\bf{W}}),
\end{equation}
where $s_c=|{\mathcal D}_c|$ is the cardinality of ${\mathcal D}_c$, $s=\sum_{c}s_c$ is the total number of data samples used in the training, $C = |\mathcal{C}|$ is the cardinality of $\mathcal{C}$, and $f_c(\cdot)$ is the local loss function of client $c$, which is given as
\begin{equation}\label{Local loss function}
f_c({\bf{W}})=\frac{1}{s_c}\sum\limits_{i\in {\mathcal D}_c}l({\bf{W}};x_i,y_i),
\end{equation}
where $l(\cdot)$ is a metric function evaluating the loss, which depends on the underlying learning model. The FL process is repeated until the model reaches a desired accuracy \cite{zhang2021adaptive}. An illustration of the federated training procedure of FL is shown in Fig.~\ref{FL}.

\begin{figure*}[!t]
\centering
\subfigure[Horizontal federated learning.]{
\includegraphics[width=2.1in]{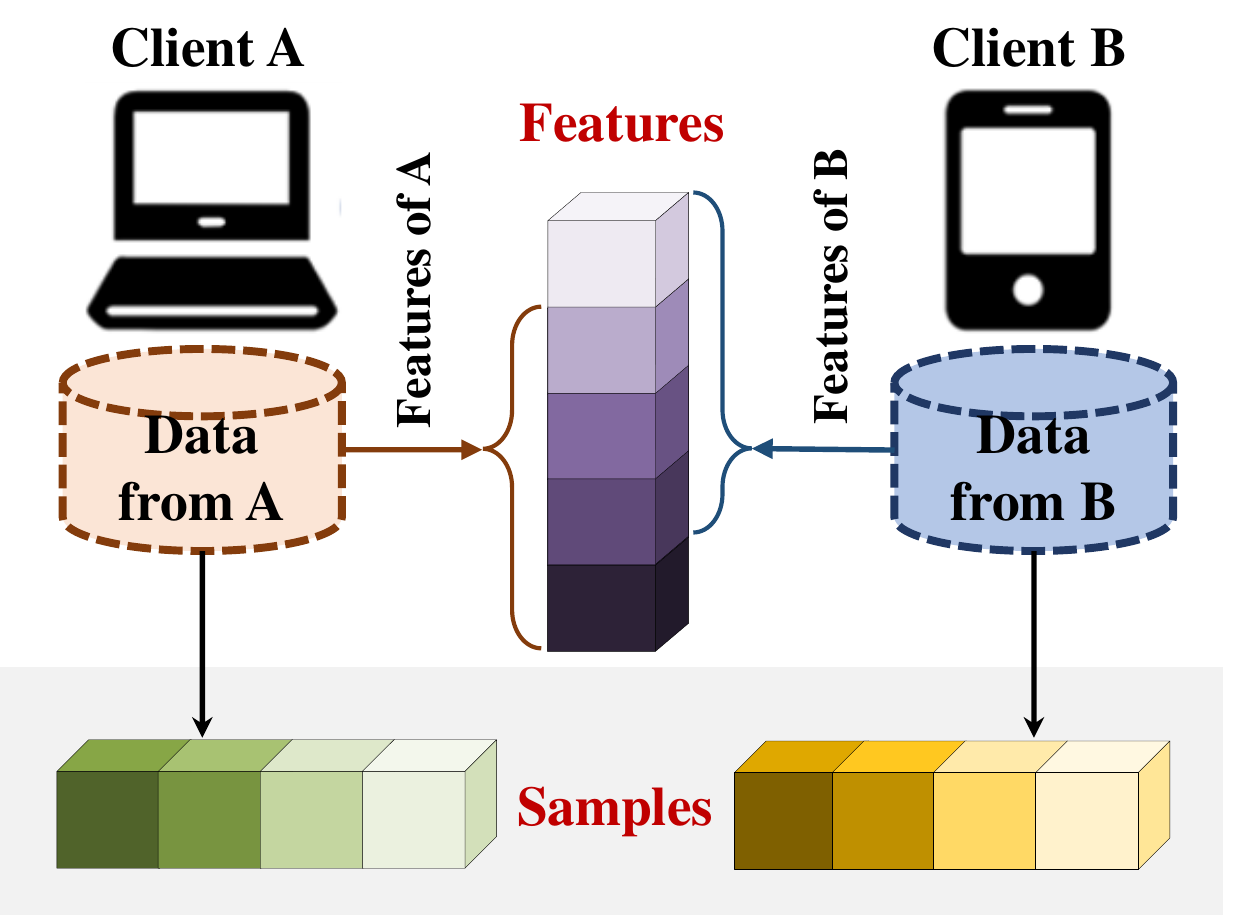}
\label{HFL}
}
\subfigure[Vertical federated learning.]{
\includegraphics[width=2.1in]{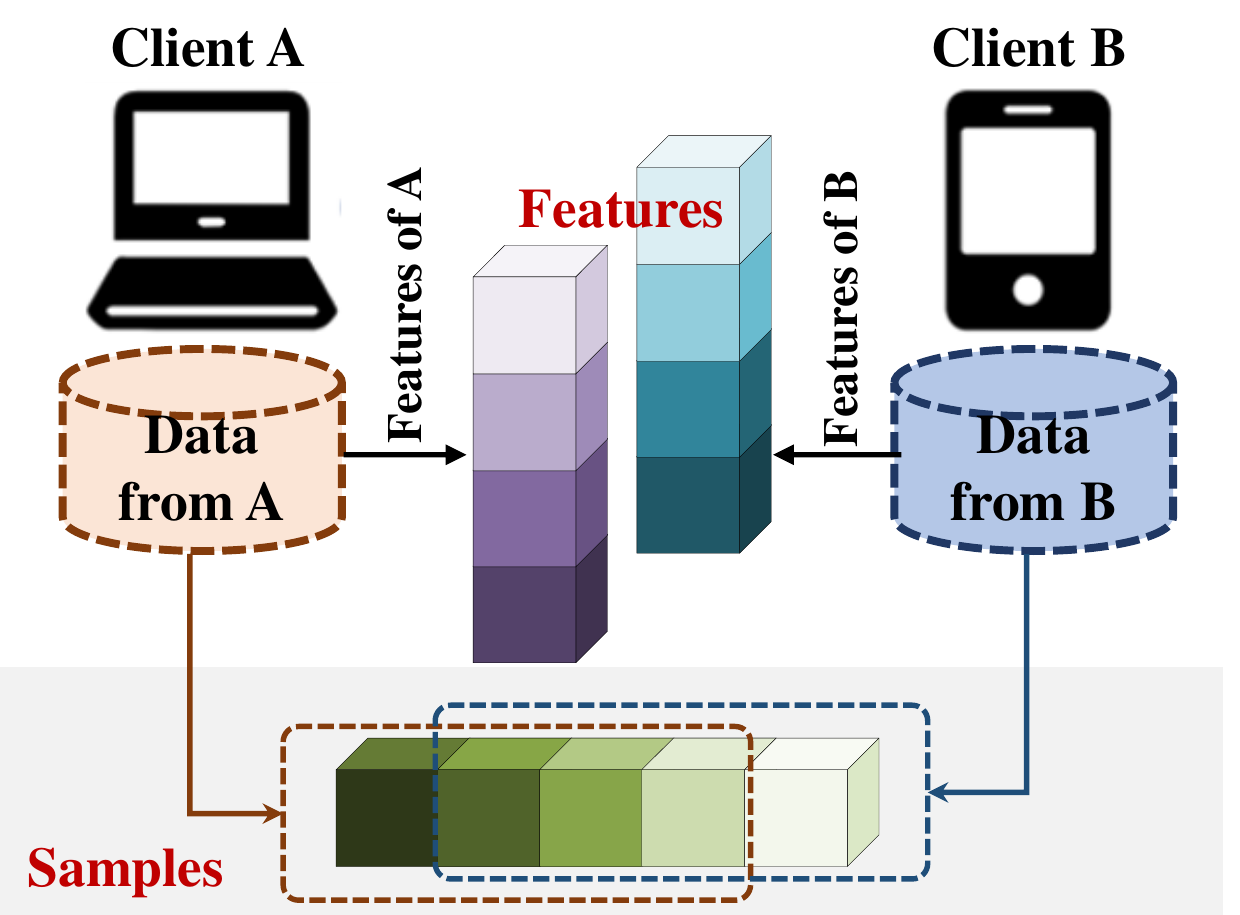}
\label{VFL}
}
\subfigure[Federated transfer learning.]{
\includegraphics[width=2.1in]{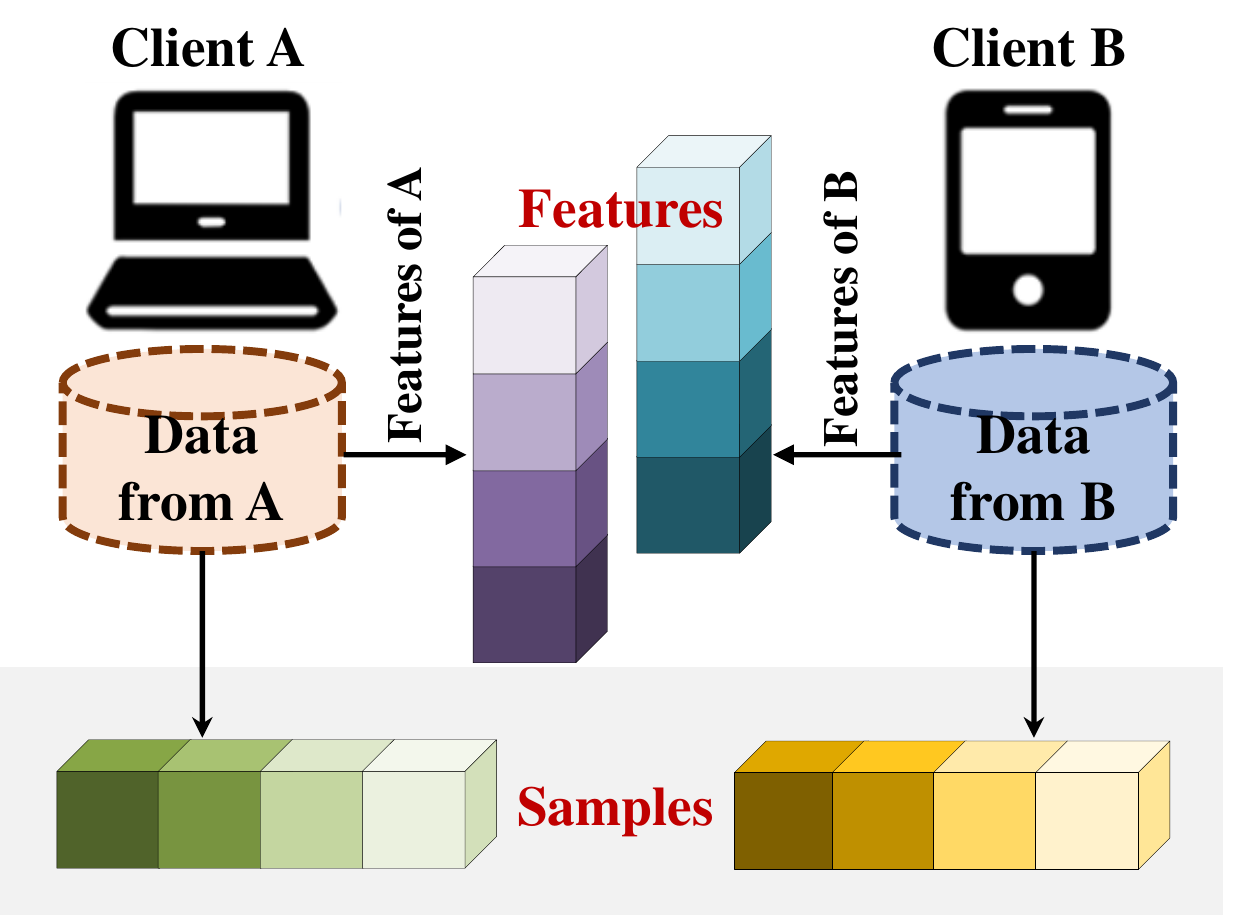}
\label{FTL}
}
\caption{Classification of FL.}
\label{FL Taxonomy}
\end{figure*}
A training dataset consists of the sample space, i.e., the data identity document (ID) space, the feature space, and the label space. According to the distribution characteristics of dataset, FL can be divided into horizontal FL (HFL), vertical FL (VFL), and federated transfer learning (FTL) as compared in Fig.~\ref{FL Taxonomy}. The HFL is a sample-based FL, where the clients share parts of a feature space, but have different sample spaces, shown in Fig.~\ref{HFL}. A typical use case of HFL is voice assistants for smart homes \cite{leroy2019federated}, in which users issue the same instruction (feature) with different types of voice (samples).
The VFL is a feature-based FL, as shown in Fig.~\ref{VFL}, where the clients share same data sample space, but have different feature spaces, e.g., regarding a user access control problem in a wireless access network, where a group of users (samples) frequently access and switch between base stations (features) \cite{cao2021user}.
Both HFL and VFL may be ineffective when the overlap of data sample space and feature space is marginal for the clients, e.g., in wearable healthcare \cite{chen2020fedhealth}. As a remedy, FTL is designed for addressing the issues in these use cases \cite{pan2009survey}. For instance, different physical characteristics and daily activity patterns (feature) of different users (sample) can be transfered to learn to develop personalized healthcare plans, where FTL applies.

The distributed architecture of FL effectively guarantees both data locality and privacy, reduces the communication cost and latency caused by data offloading, and provides high learning quality. Specifically, FL allows devices to collaboratively train a global model without sharing personal data. Different from collecting all data to train a model by centralized ML methods, FL meets the requirement for data privacy and security provision \cite{gu2021privacy}. For example, in \cite{domingo2021secure}, FL was applied to guarantee privacy protection and security resistance to participating devices. Furthermore, large offloading latency is avoided in FL, since it does not need to offload raw data to the central server. In particular, for edge devices with insufficient computing power, the distributed training of FL significantly reduces model training latency \cite{yang2019federated,nishio2019client}. Additionally, by collecting large and diverse datasets from many devices, FL also improves the convergence rate of training and obtains an accurate global model \cite{nguyen2021federated}. As such, edge devices with insufficient local data also benefit from the collaborative training of FL.

Despite its various advantages, personalization-related FL systems raise the following unique challenges that are different from the cloud data center-based learning model \cite{li2020federated}. A growing body of recent researches, e.g., \cite{mcmahan2017communication,konevcny2016federated,xia2021federated,wu2020personalized,smith2017federated,caldas2018federated,xu2020fedmax,gu2021privacy,zhao2020local,ding2019novel}, have developed effective methods to deal with these challenges.
\begin{itemize}
\item \emph{Communication cost}: The distributed training architecture of FL comes with frequent exchanges of model parameters between the central server and clients, resulting in high communication cost. Model compression can be used to reduce the handover load in each communication round \cite{konevcny2016federated}, while the required number of communication rounds can be minimized by using techniques such as the federated average approach in \cite{mcmahan2017communication}. Exchange of only important gradients, the importance-based updating in \cite{xia2021federated}, is another potential approach for the cost reduction.

\item \emph{Statistical heterogeneity}: In wireless edge networks, it is often unrealistic to assume that edge devices generate i.i.d. datasets of similar sizes. Often local data do not follow the same distribution as that of the overall data \cite{verma2019approaches}. In practice, cross-device collaborative learning architectures of FL with non-i.i.d. local data leads to statistical heterogeneity. In \cite{wu2020personalized}, it was found helpful to address the statistical heterogeneity by embedding the notion of personalization in FL to capture non-specific aspects. Also, in \cite{smith2017federated,caldas2018federated}, a multi-task learning framework was proposed to address the heterogeneity challenge.

\item \emph{System heterogeneity}: FL in IoE systems often involves numerous devices, such as smart phones, laptops, and wearable devices, with different computing power, storage capacity, and battery lifetime. Since the update efficiency of gradient update per training round is determined by the device with the most constrained capabilities \cite{cui2021client}, it is therefore inefficient, sometimes even intractable, to consider all clients in each update round. This synchronously distributed training pattern of FL leads to the challenge referred to as system heterogeneity \cite{nguyen2021federated1}. As part of the solution, a subset of clients are randomly or deterministically scheduled to perform distributed training per update round \cite{mcmahan2017communication,xu2020fedmax}.

\item \emph{Privacy concern}: Although in FL nodes do not reveal their local data to the others, there still are security and privacy vulnerabilities at both the central server and clients. This issue may prevent widespread adoption of FL in many wireless IoE applications, e.g., vehicle-to-vehicle (V2V) communication, healthcare, and smart home. Recent studies have demonstrated that the process of model sharing and update in FL poses a potential threat of information leakage and privacy violations \cite{melis2019exploiting}. Furthermore, malicious attackers can infer individual clients' private information of clients by observing the transmitted gradients. In order to protect privacy, a secure multiparty computation algorithm was proposed for FL in \cite{gu2021privacy}. Also, in \cite{zhao2020local,ding2019novel}, the addition of noise to raw data and the use of differential privacy (DP) methods were shown effective in privacy protection.
\end{itemize}

\subsubsection{Split Learning}
Unlike FL, where clients and the server need to train a full ML model, split learning (SL) is another distributed ML method, where the clients and server only need to train a part of the entire model. In SL, neither raw data nor the model architecture and weights are shared among clients and the server such that they cannot access other's models\cite{gupta2018distributed}. Concrete differences between FL and SL are compared in Table~\ref{Comparison}.

The crux of SL is to split the entire neural network into parts and deploy the split parts on clients and server respectively. Each client device retains a part of the neural network, and the network parts of all devices constitute a complete model \cite{gupta2018distributed}. Importantly, the splitting strategy significantly affects the learning performance. In general, there are three levels of network splitting for SL\cite{vepakomma2018split}. A basic process of SL includes splitting the network and training. The network is first split into two parts. The first part, denoted by $\mathcal{N}_c$, lies in a client, and the other part, denoted by $\mathcal{N}_s$, is located on the server. There is a boundary layer between the two parts, called a cut layer. The client inputs the source data into $\mathcal{N}_c$ to execute forward propagation and outputs $\mathnormal{C}_{\rm{out}}$ at the cut layer. The output $\mathnormal{C}_{\rm{out}}$ and the label are sent to the server as the input of $\mathcal{N}_s$ to obtain the output. Gradients are calculated using the transmitted labels and are backpropagated to the terminal client. These steps are repeated until the model converges. An extension of the basic SL is to networks with multiple clients, where each client has a different partial network that produces different outputs at the cut layer. The gradients are calculated and backpropagated in the same way as the basic SL, and multiple clients can cooperate to complete the target task without sharing the raw data. Both of the basic SL and extended SL methods need the clients and server to share labels. A configuration that does not require label sharing, called U-shaped SL configuration, was proposed in \cite{vepakomma2018split}.

SL enjoys many advantages over traditional DL methods. Especially for applications in wireless communications, the vigorous development of IoE has caused a surge in the number of mobile devices generating massive data. Due to limited computing power of most IoE devices, we usually integrate all data to the server for centralized ML, which however causes potential information leakage and increases processing delay. In SL, the server is prevented from accessing client's networks and data, which protects privacy to some certain extent. Meanwhile, SL distributes the training tasks and thus eases computational burden on clients. In addition, SL does not share the raw data, thus reducing the communication bandwidth required for information exchange. In \cite{gao2020end}, it has been experimentally verified that SL achieves better accuracy and faster convergence than FL when data distributions at multiple clients are imbalanced.

Thanks to these advantages, SL has been used in wireless networks for millimeter-Wave (mmWave) communications \cite{koda2020communication}\cite{koda2020distributed}, unmanned aerial vehicle (UAV) networks \cite{liu2022energy}, mobile edge networks \cite{tian2021jmsnas}, etc. In \cite{koda2020communication}, a distributed multimodal ML framework, called multimodal split learning (MultSL), was proposed to improve the accuracy of mmWave received power prediction while protecting privacy. In this framework, a convolutional long short-term memory (LSTM) neural network is split into two segments which are, respectively, deployed in the user equipment (UE) and the base station (BS). The UE, with a camera collecting images, extracts image features through the partial neural network. The RF signal received by the BS is processed by the other partial network on its side. The features are combined at the BS to predict the receive power. Since the methodology does not make use of raw images and RF signals, this SL method boosts privacy. This approach was then extended in \cite{koda2020distributed} to multiple UE cameras. The authors proposed heteromodal SL with feature aggregation, which improved the method in \cite{koda2020communication} in terms of both accuracy and privacy. Also, in \cite{liu2022energy}, a hybrid split and federated learning (HSFL) framework was proposed for data analysis and inference in UAV networks. The scheduled UAVs select SL or FL training methods according to their computing powers. The UAV and the BS cooperatively train a part of the DNN when the UAV chooses the SL method. It turns out that HSFL reduces energy consumption compared to FL and split federated learning (SFL) methods while preserving accuracy. The idea of SL was also used to segment DNNs in mobile edge networks and a joint model split and neural architecture search framework was developed in \cite{tian2021jmsnas}. This framework uses neural architecture search method to split the DNN in the edge mobile computing (MEC) according to the computing power and communication capacity of MEC device. The results showed that this splitting method achieves higher accuracy and lower latency than the state-of-the-art methods such as MobileNet \cite{9008835} and HiveMind \cite{9562523} multi-split frameworks.

\subsubsection{Multi-agent Reinforcement Learning}
As a central ML paradigm, RL \cite{sutton2018reinforcement} has contributed enormously to the development of AI in recent years. Specifically, the single-agent RL is mainly used to solve sequential decision problems, which are generally modeled as Markov decision processes (MDP). Combining RL with DNN, deep reinforcement learning (DRL), e.g., deep Q-learning (DQL)\cite{mnih2013playing}, has emerged as a powerful tool to solve resource allocation problems in many wireless applications, e.g.,~\cite{9120241,8796358,8633948,9046301}.

In single-agent RL/DRL, an agent centrally processes all information from environment. However, various emerging services, such as MEC, IoE, and the industrial Internet, causes the number of user equipments to grow. The B5G networks, developing in a decentralized, self-organizing, and autonomous, are expected to serve massive connected devices with ultra reliability and low latency. Single-agent RL approaches are no longer suitable to meet these challenging requirements. To address these challenges, Multi-agent RL (MARL) generalizes the single-agent RL to settings with multiple controllers. MARL consists of a set of physically or logically distributed agents that can interact not only with the environment but also other agents to acquire optimal policies\cite{busoniu2008comprehensive}.

\begin{figure}[!t]
\centering
\includegraphics[width=3.3in]{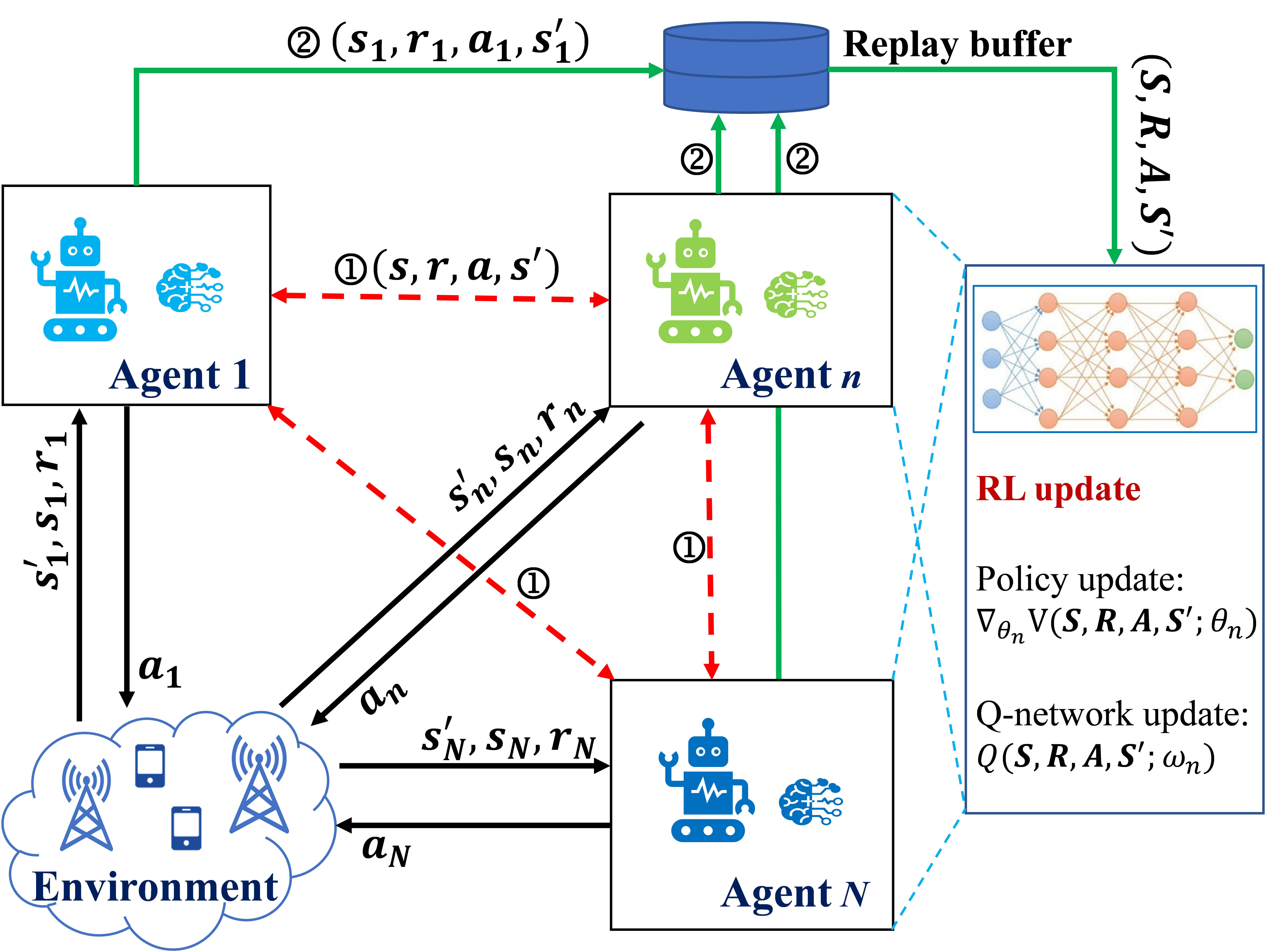}
\caption{Interaction between environment and agents in distributed MARL.}
\label{fig-MARL-interaction}
\end{figure}

Unlike single-agent RL, MARL is usually modeled as a Markov game (MG) or stochastic game (SG) \cite{shapley1953stochastic}. Specifically, an SG can be defined by a tuple $\left \langle  \mathcal{N,S,A},P,\mathcal{R},\gamma \right \rangle$, where $\mathcal{N}$ is the set of agents, $\mathcal{S}$ is the set of state spaces of all agents, $\mathcal{A}= \mathcal{A}_1\times...\times \mathcal{A}_N$ is the joint set of action space, $\mathcal{A}_n$ is the action space of agent $n,n \in \mathcal{N}$, $P$ represents the transition probability function from the current state $\tilde{\mathcal{S}}$ $\in$ $\mathcal{S}$ to the next state $\tilde{\mathcal{S}}^\prime$, $\mathcal{R}= \{r_1,...,r_N\}$ is the set of reward functions of all agents which depends on their actions, and $\gamma \in [0,1)$ denotes a discount factor. Interactions between environment and agents in distributed MARL are illustrated in Fig.~\ref{fig-MARL-interaction}.

In each discrete time step $t$ in MARL, every agent $i$ selects an action $a_{i,t}$ based on the current state $\tilde{\mathcal{S}}$, and receives an immediate reward $r_{i,t}$. The environment state transits to the next state according to the action set $\tilde{\mathcal{A}} =\{ a_{i,t}, \ i \in \mathcal{N} \}$. Agent $i$ aims to find its optimal policy $\pi_{i}^*$ to maximize its own discounted accumulative reward. This policy, however, depends on the joint policy $\pi=\prod_{i\in \mathcal{N} }\pi_{i}$ of all agents. To determine agent actions, two important functions, i.e., a state-value function and an action-value function, a.k.a. Q-value function, are defined for each agent $i$ as follows:

\begin{align}
Q_{\pi_{i}}(\tilde{\mathcal{S}},\tilde{\mathcal{A}}) &= \mathbb{E} \left\{ \sum_{j=0}^{\infty} \gamma^j r_{i,t+j}|\tilde{\mathcal{S}},\tilde{\mathcal{A}},{\pi}\right\}, \quad i\in\mathcal{N},\\
V_{\pi_{i}}(\tilde{\mathcal{S}}) &= \mathbb{E} \left\{ \sum_{j=0}^{\infty} \gamma^j r_{i,t+j}|\tilde{\mathcal{S}},{\pi}\right\}, \quad i\in\mathcal{N},
\end{align}
where $r_{i,t+j}$ is the reward of $j$ steps after time step $t$ of agent $i$ and $\mathbb{E}\{\cdot\}$ takes the average of the long-term discounted rewards.

In distributed MARL, each agent updates its own policy locally, but this process requires information from other agents. As shown in Fig. \ref{fig-MARL-interaction}, the agents obtain the information by interacting with other agents or from a replay buffer that stores the information.
Then, MARL involves the interaction among multiple agents whose rewards not only depend on their own states, but also are affected by the other agents. A comprehensive and reasonably designed reward function plays a crucial role in solving these problems. According to the types of reward functions, MARL algorithms are classified into three categories: \textit{fully cooperative, fully competitive, and mixed} MARL\cite{bucsoniu2010multi}.

In fully cooperative MARL algorithms, all agents share the same reward function, i.e., $R_1=...=R_N=R$. Agents cooperate with each other to achieve the same goal. Based on the amount of information shared between agents, there are two types, i.e., independent MARL and collaborative MARL \cite{chen2021distributed}. For independent MARL, the agents have access to their own local information and optimize their policies independently. For example in \cite{cui2019multi}, an Independent Learner (IL) MARL algorithm was proposed to solve a dynamic resource allocation problem in a multi-UAV network, where the quality of service (QoS) is defined as the reward function and each UAV is an independent agent with only local channel state information. As for collaborative MARL, the agents can share,
at least partially, information with each other. This kind of MARL was used for trajectory design in UAV networks \cite{9209079} and task offloading in MEC \cite{9037194}.

In fully competitive MARL algorithms, multiple agents have conflicting goals, and each agent desires to maximize its own reward while minimizing the opponents' reward. It is often defined as zero-sum MGs, i.e., $\sum_{i=0}R_i=0$. A typical algorithm is Minimax-Q \cite{littman1994markov}. Note that this kind of algorithm is applicable to scenarios with competitive players. For example, jamming attack in a cognitive radio network often uses this algorithm to maximize the spectral efficiency \cite{5738229}, where secondary users and attackers are modeled as two opposite players with opposite reward functions.

Mixed MARL algorithms combine the characteristics of cooperation and competition. There is no clear restriction on the relationship between the reward functions of agents. It is generally defined as a general-sum game. Algorithms of this type include Nash Q-learning\cite{hu2003nash}, correlated Q-learning, etc. In wireless networks, they are often used in heterogeneous networks. For example, a network selection algorithm based on Nash Q-learning was proposed in\cite{9403383} for an heterogeneous network where different types of networks are the agents striving to provide service for users with different requirements. Reward functions of these agents were defined by network utilities with different expressions, depending on their serving users.

MARL enjoys many advantages compared to single-agent RL. Multiple agents can solve problems in a distributed and parallel manner, which improves the efficiency of the algorithm. Moreover, MARL is more scalable and robust compared to single-agent RL. As each agent learns its own policy, sporadic changes in the number of agents has little impact on the policy learning process of other agents.

Although MARL has made considerable progress especially in EL, there are still many challenges to be addressed toward its deployment in real-world applications.
\begin{itemize}
\item \emph{Non-stationary environment}: In a multi-agent system, agents learn their policies simultaneously. Each agent has to jointly consider both the actions of the other agents and its own action. These interactions with other agents constantly alter the environment, which makes it difficult for all agents to obtain their optimal policies. Considering the distributed implementation of MARL, a frequently adopted solution is centralized training and distributed execution (CTDE). For example, a CTDE method was used in \cite{9120241} to optimize the power allocation in a multiuser cellular network with MARL.

\item \emph{Partial observation}: In practice, an individual agent usually has access to partial state information, which impairs their ability to learn the globally optimal strategy. In \cite{9277917}, it was shown that a consensus communication mechanism with a graph network-based self-attention can effectively reduce the effect of partial observation on MARL in a dynamic environment with device-to-device (D2D) communications.

\item \emph{Training approach}: Many multi-agent algorithms exploit a fully centralized or fully distributed training approach. In the fully centralized approach, a central unit is responsible for policy learning with data from all agents, as shown in Table~\ref{Comparison}. This approach suffers from high computational complexity. However, fully distributed training approaches suffer from convergence issues due to the availability of only partial state information for training. The approach of CTDE \cite{calabrese2018learning} has been proven to be more effective than fully centralized and fully distributed training modes. With CTDE, a centralized network uses global information for centralized training, and the learned policy is distributedly executed by agents with their own local information. It alleviates the problems caused by non-stationary environments, ensures convergence, and reduces training overhead.

\end{itemize}

\subsection{Inference of Edge Learning}
Along with the above distributed model training methods, inference is another important component procedure of ML by applying a pre-trained model to new data and making a decision or prediction. Due to the concerns of data privacy, latency, energy consumption, and unstable network connection, centralized inference~(CI) at a data center can hardly satisfy these demanding requirements of massive edge devices. On the other hand, executing inference locally on edge devices requires significant computational resources, which is often impractical in IoE.

To address these problems, some recent researches have focused on applying various techniques, e.g., sparsification, and pruning techniques \cite{bhattacharya2016sparsification,jabbar2020mri,li2018edge}, to enable distributed inference (DI) with improved efficiency and performance. In \cite{jabbar2020mri}, a fuzzy DI technique was developed to recognize objects in remote medical imaging videos, which obtains acceptable inference accuracy with extremely low latency. In \cite{li2018edge}, an on-demand DI framework was designed for  edge devices to conduct collaborative inference.

In ML, the stage of model training has been considered as the most computationally intensive stage. Although computational requirements for inference are typically lower than that for training, it is still a huge challenge for edge devices with insufficient computing capabilities to perform DI frequently \cite{wu2019machine}. On the other hand, as edge devices are highly heterogeneous in terms of hardware specifications and usage scenarios, there does not exist a universal model that fits edge devices from all aspects, e.g., accuracy, latency, and energy consumption. To tackle these challenges, a once-for-all network was proposed in \cite{cai2019once} to determine the inference model. It surprisingly fits different hardware conditions and latency constraints. Alternatively in \cite{lu2019automating}, an automated DNN model selection algorithm was developed for DI, which highlights the potential of learning model selection.

%% file: tables/table3.tex
\begin{table*}[htbp]
	\centering
	\caption{Comparison of Centralized Learning and Distributed EL}
	\resizebox{\linewidth}{!}{
		\begin{tabular}{c|c|c|c|c|c|l|l}
			\toprule
	         \multirow{2}{*}{\textbf{ML phase}} & \multirow{2}{*}{\textbf{Method}} & \multirow{2}{*}{\textbf{Topology}} & \multicolumn{1}{c|}{\textbf{Cloud} } & \textbf{Local client}  & \multicolumn{1}{c|}{\textbf{Exchanging} } & \multicolumn{1}{c|}{\multirow{2}{*}{\textbf{Pros.}}} & \multicolumn{1}{c}{\multirow{2}{*}{\textbf{Cons.}}}\\
             &       &       & \multicolumn{1}{c|}{\textbf{server}} & \textbf{(or agent)} & \multicolumn{1}{c|}{\textbf{information}} &  &\\
			\midrule
			\cline{1-8}
			\multirow{8}{*}{\textbf{Trainning}} & \multirow{2}{*}{Client-server} & \multirow{2}{*}{Centralized} & \multirow{2}{*}{Model \& data} & \multicolumn{1}{c|}{\multirow{2}{*}{Data}} & \multirow{2}{*}{Data} & Easy to develop and  &  Long delay and no \\
			&       &       &  & \multicolumn{1}{c|}{} &       &   maintain & privacy guarantee \\
			\cline{2-8}
			& \multirow{2}{*}{FL} &       & \multirow{2}{*}{Model } & \multirow{2}{*}{Model \& data} & Model \& model  & Low offloading cost & Computing idle in \\
			&       & \multicolumn{1}{c|}{Centralized/} &       &   & \multicolumn{1}{c|}{updates} &  and latency  & model synchronization \\
			\cline{2-2}\cline{4-8}
			& \multirow{2}{*}{SL} & \multicolumn{1}{c|}{Decentralized/} & \multirow{2}{*}{Partial model} & Partial model  & \multicolumn{1}{c|}{Forward tensors \& } & Privacy protection, low &  Hard to design\\ 
			&       & \multicolumn{1}{c|}{Distributed} & & \& data & \multicolumn{1}{c|}{backward gradients} & commun. bandwidth & and slow training \\
			\cline{2-2}\cline{4-8}
			& \multirow{2}{*}{MARL} &       & Model \& data &\multirow{2}{*}{Model \& data}  & \multicolumn{1}{c|}{Data (state,} & Adaptive to changing  &  \multirow{2}{*}{Hard to converge}\\ \cline{4-4}
			&       &       & N/A      &  & \multicolumn{1}{c|}{   action, reward)} &  environments & ~\\ 
			\midrule
			\cline{1-8}
			\multirow{4}{*}{\textbf{Inference} } & \multirow{2}{*}{CI}    & \multirow{2}{*}{Centralized} & \multirow{2}{*}{Model} & \multirow{2}{*}{Data} & \multirow{2}{*}{Data}  & No computing power & Long delay and no \\
			&       &       &       &   &  & requirement for clients   & privacy guarantee \\
			\cline{2-8}
			& \multirow{2}{*}{DI} & \multirow{2}{*}{Distribued} & \multirow{2}{*}{N/A} & \multirow{2}{*}{Model \& data}  & \multirow{2}{*}{N/A} & Privacy protection and & Additional resource   \\
			&       &       &       &   &   & low latency  & requirement at clients \\
			\bottomrule
		\end{tabular}%
	}
	\label{Comparison}%
\end{table*}%

%% file: sections/section2-C.tex
\subsection{Joint Edge Learning and Communication Resource Allocation Optimization}

\begin{figure*}[t]
\centering
\includegraphics[width=7.0in]{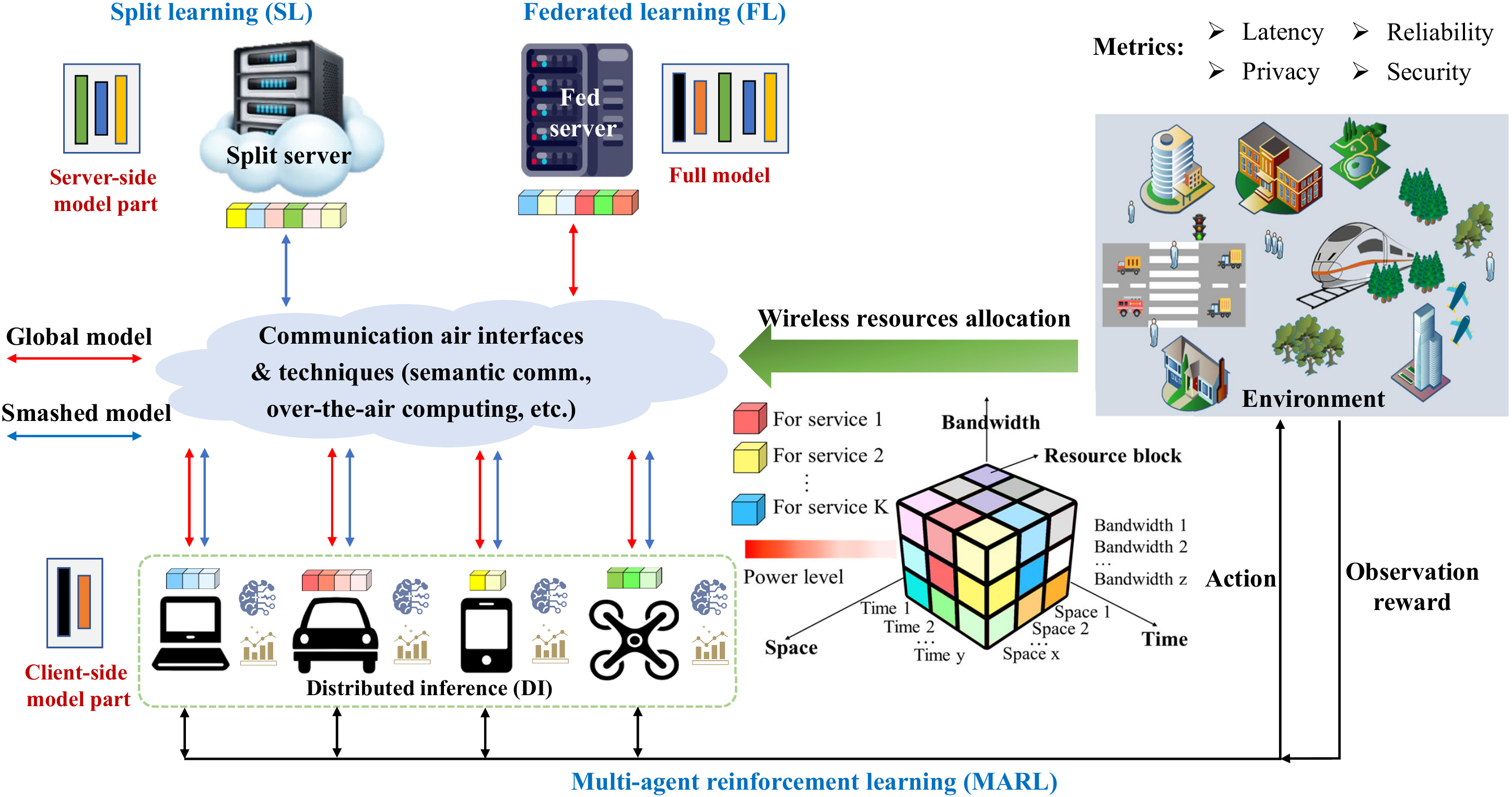}
\caption{An illustration of joint EL and communication ecosystem.} \label{fig4}
\end{figure*}
As we alluded to, there exists a symbiotic relationship between EL and wireless communications. On the one hand, EL plays a critical role in optimizing link performance in wireless communication systems. On the other hand, the functionalities and performance of EL depend highly on communication ability, especially when compute nodes are connected by wireless channels. To be specific, the design of EL architectures and their operations should be jointly optimized under communication and on-device resource constraints \cite{lin2020distributed}. Besides popular concerns of latency and reliability, additional aspects should be counted in the on-device constraints, e.g., energy, computation, caching memory, and privacy. From a theoretical standpoint, a joint learning and communication optimization paradigm provides a unified framework to fully utilize communication theory, offer fundamental privacy and security guarantees, and reap promised performance gains for ML at network edges \cite{park2019wireless}. Although studies on this exciting new area is in its infancy, preliminary efforts have been devoted to fully explore the key building blocks, principles, and applications of EL, as well as their connections with distributed wireless communication. We review some state-of-the-art literature on FL, RL, SL, and DI from a joint communication and learning standpoint. In Fig. 6, the interplay and joint optimization of resource allocation for wireless communication and these EL techniques are illustrated.

Taking FL as an example, the trade-off between learning time and UE energy consumption and the trade-off between computation time and communication latency are of wide interest. As a first attempt, \cite{zeng2020energy} considered an energy-efficient resource allocation strategy for FL by bandwidth allocation and scheduling. In \cite{mo2021energy}, by taking into account both communication resources and computing power for learning, the energy consumption at all edge devices is minimized for training. In addition, to alleviate the ``straggler effect” where the slowest edge device acts as a bottleneck of learning performance \cite{ha2019coded}, a new protocol for FL was advocated in \cite{nishio2019client} through joint optimization of heterogeneous data, computing power, and communication resources, where only the edge devices with good communication and computation qualities are chosen.

As for RL, it has been widely used for learning to solve resource allocation problems in wireless communication systems. In \cite{wang2021a}, a multi-stack RL method was proposed for task and resource allocation in MEC. Also in \cite{anh2019efficient}, a DRL algorithm was devised for efficient training management which exhibited superiority in terms of both energy consumption and training latency. For a UAV-communication system, a DRL-based collaborative optimization was developed in \cite{huang2021deep} for communication resource allocation and UAV route planning, achieving real-time obstacle avoidance. 

When it comes to SL and DI, some recent studies have started investigating their joint optimization with communications. It has been shown in \cite{Gunduz2020communicate} and \cite{jankowski2020deep} that separate communication and inference design at the network edge can be highly suboptimal. In contrast, a joint optimization of communication and inference helps improve both the accuracy and speed of inference. Inspired by this, a communication-efficient SL framework was proposed in \cite{krouka2021communication} to cope with the problems of limited bandwidth and noisy time-varying channels. While in \cite{koda2020communication}, SL is considered for received power prediction for mmWave systems in a privacy-preserving manner. 

The optimization of communication resources is of significant importance in improving the performance of EL in not only training but also inference stages. In what follows, we will focus on distributed optimization techniques to enable service-driven resource allocation in B5G networks under the aforementioned communication and on-device constraints.

%% file: sections/section3-A.tex
\section{Edge Learning and Communication Optimization}

In this section, we investigate the interplay between EL techniques and wireless communication resource allocation optimization.
We first introduce the performance metrics for EL network design, capturing both learning and communication performance. 
Thereafter, we provide a holistic overview of key optimization methods for EL. 
Finally, we end this section with discussions on the convergence and signaling overhead of these algorithms and methods.

\subsection{Dual-functional Performance Metrics for Learning}

It is expected that edge networks will serve as a key enabler for the future 6G intelligent networks. 
Compared to conventional centralized ML approaches that require a central controller to train large datasets, each user terminal in EL collects its local data for the neural network training without raw data exchanged among users, as such data is usually private or cannot be transmitted completely due to limited wireless resource.
In EL, all terminals need to train a common global model collaboratively. Implementing EL networks usually relies on data and gradient signaling over wireless channels, which inevitably suffers from transmission error and delay due to channel fading and interference. 
To mitigate interference and cope with fading effects, it is important to investigate optimizing EL design over wireless channels with limited resources for communications.
Therefore, various schemes have been proposed to solve these difficulties in EL under wireless communication constraints, aiming to improve learning performance in terms of model accuracy, convergence, privacy protection, and network security, as summarized in Table III.

\subsubsection{Accuracy and Convergence}

Similar to traditional ML networks, model accuracy and convergence are important aspects of EL optimization. 
Especially in edge networks, edge terminals can only share a part of the processed data with each other as well as the central servers through wireless channels, which makes it challenging to guarantee learning accuracy and convergence of EL methods with imperfect and outdated signaling.

A typical edge network accuracy optimization problem can be formulated as
\begin{equation}
	\underset{{\bm \theta} \in \mathbb{R}^d}{\rm minimize}\quad \mathcal{L}({\bm \theta}):= \sum_{k \in \mathcal{S}}{w}_k\mathcal{L}_k({\bm \theta;\mathcal{D}_k}),
\end{equation}
where ${\bm \theta}$ is used for representing the real-valued network parameters, $\mathcal{L}_k(\cdot)$ is the local loss function of the $k$-th terminal calculated using the dataset $\mathcal{D}_k$, $\mathcal{S}$ denotes the set of all edge terminals, and $w_k \ge  0$  is the weight for each local loss function with $\sum_{k \in \mathcal{S}} w_k = 1$ \cite{letaief2021edge}.
For edge inference, network accuracy is defined as the completion quality of the given task, which is closely related to the target task, training dataset, and communication quality. 
Based on the accuracy requirements of an edge inference network, the authors of \cite{letaief2021edge} constructed a unified framework for service resource allocation in the edge network and presented optimization algorithms based on mathematical programming and ML. 
Besides, the authors of \cite{shao2021learning} tried to reduce the consumption of communication resources in edge reasoning and improve the accuracy of the network by using the guideline of information bottleneck, i.e., maximizing the mutual information between the inference result and the coded feature, while minimizing the mutual information between the coded feature and the input data.

Different from edge inference, in FL, the quality of iterative updates of the global model depends on the gradient information fed back from edge devices over a wireless network.
Due to limited transmission bandwidth, selection of user gradient information for aggregation is crucial \cite{chen2020joint}. 
A jointly optimized resource block allocation, power control, and user selection for FL was proposed to address this issue. 
Alternatively, in \cite{liu2021hierarchical}, a new framework was proposed by letting models of edge users be quantized and aggregated in the edge server before being uploaded to the central server for global model aggregation. 
In addition, the convergence bound of this new framework was deduced, and the impact of quantization of shared parameters on convergence of the model was also investigated.
Then in \cite{magnusson2020maintaining}, an adaptive quantizer was proposed, which shows that transferring only a few bits per iteration is sufficient for ensuring linear convergence.

\subsubsection{Privacy Protection}


Privacy protection is a key indicator in future intelligent communication systems. 
Traditional centralized networks require raw data that contains the private information collected from users for model training. In edge networks, the original user data is not transmitted between edge terminals and central servers, as the user data privacy is protected by methods like differential privacy and FL framework.

Differential privacy is a common method used to protect user privacy. A randomized function $ \mathcal{M}(\cdot)$ gives $\epsilon$-differential privacy, if for all datasets $\mathcal{D}_1$ and $\mathcal{D}_2$ differing on at most one element and
\begin{equation}
	\frac{P(\mathcal{M}(\mathcal{D})\in \mathcal{S})}{P(\mathcal{M}(\mathcal{D'})\in \mathcal{S})} \leq e^{\epsilon},
\end{equation}
where $\epsilon$ is a small positive number and $P(\cdot)$ is the probability of an occurrence. The notation $\mathcal{M}(\mathcal{D})\in \mathcal{S}$  means that the output of algorithm $ \mathcal{M}(\cdot)$ with inputting $\mathcal{D}$ is in the range of $\mathcal{S}$ \cite{dwork2008differential}.

A randomized mechanism ${M}$ space $:{X}^{K\times m}\rightarrow{\mathcal{R}}^d$ is said to be $\rho$-zero-concentrated differentially private (CDP) if for every adjacent of ${X}$, denoted by $\tilde{{X}}\in{X}^{K\times m}$, it holds that
\begin{equation}
	{D}_{\alpha}\big({M}({X})\,\|\,{M}(\tilde{{X}})\big)\leq \rho\alpha,
\end{equation}
for all $\alpha\in(1,\infty)$, where ${D}_{\alpha}$ is the $\alpha$-R\'enyi divergence \cite{bun2016concentrated}.

\begin{table*}[]
	\caption{Dual-functional Performance Metrics for Learning and Communication}
	\centering
	\setlength{\tabcolsep}{2.3mm}{
		\begin{tabular}{|c|c|c|l|c|}
			\hline
			\begin{tabular}[c]{@{}c@{}}\textbf{Dual-functional}\\ \textbf{Performance}\end{tabular}                      & \textbf{Performance Metrics}                                                                                 & \textbf{References}     & \multicolumn{1}{c|}{\textbf{Objective}}                                                                                                                                                                                                       & \textbf{Method }                \\ \hline
			\multirow{9}{*}{\textbf{\begin{tabular}[c]{@{}c@{}}Learning\end{tabular}}}       &\multirow{3}{*}{\begin{tabular}[c]{@{}l@{}}\\Accuracy and\\ convergence\\ \end{tabular}}&{\cite{letaief2021edge}}, {\cite{shao2021learning}} & \begin{tabular}[c]{@{}l@{}}Resource allocation for improving accuracy in training of EL\end{tabular}  & \begin{tabular}[c]{@{}l@{}} Optimization\end{tabular}\\ \cline{3-5} 
			&  &   {\cite{chen2020joint}},{\cite{liu2021hierarchical}}  &\begin{tabular}[c]{@{}l@{}} Accuracy and latency tradeoff in EL with convergence analysis \end{tabular} & Optimization \\ \cline{3-5}
			&  &   {\cite{magnusson2020maintaining}} & \begin{tabular}[c]{@{}l@{}} Data communication compression for EL with guaranteed linear\\ convergence \end{tabular}  & Optimization \\ \cline{2-5}
			
			& \multirow{2}{*}{\begin{tabular}[c]{@{}l@{}}\\Privacy protection\end{tabular}}&{\cite{du2020approximate}} & \begin{tabular}[c]{@{}l@{}} Privacy protection in FL via power control \end{tabular}  & DL \\  \cline{3-5} 
			& &  {\cite{arachchige2019local}}, {\cite{liu2020privacy}} & \begin{tabular}[c]{@{}l@{}} Privacy protection and efficient communication in FL \end{tabular} & DL \\ \cline{3-5}
			& & {\cite{tabassum2021privacy}} & \begin{tabular}[c]{@{}l@{}} Privacy protection by incremental learning in distributed networks\end{tabular} & DL \\ \cline{2-5}
			
			& \multirow{2}{*}{\begin{tabular}[c]{@{}l@{}}Security\end{tabular}}  & {\cite{chaabouni2019network}}, {\cite{sthapit2021reinforcement}}& \begin{tabular}[c]{@{}l@{}} Network intrusion detection using DL \end{tabular}  & DL  \\ \cline{3-5}
			& & {\cite{ferrag2021federated}}& \begin{tabular}[c]{@{}l@{}} Achieving cyber security using FL  \end{tabular}  & FL  \\ \cline{3-5}
			\hline

			\multirow{9}{*}{\textbf{\begin{tabular}[c]{@{}c@{}}\\ \\ \\ \\ \\ \\ \\  Communication \end{tabular}}} 
			& \multirow{8}{*}{\begin{tabular}[c]{@{}c@{}} Spectral efficiency \\ optimization\end{tabular}} & {\cite{park2019distributed}} & \begin{tabular}[c]{@{}l@{}} A framework of communication resource share and reallocation in a \\ self-adaption manner \end{tabular}& Optimization \\ \cline{3-5} 
			&  & {\cite{wang2019joint}} & \begin{tabular}[c]{@{}l@{}} A three-layer network of cloud-and-edge nodes for computing and c- \\ ommunication resource reuse  \end{tabular} & Optimization \\ \cline{3-5} 
			&  & \begin{tabular}[c]{@{}c@{}}{\cite{chen2018label}}, {\cite{alqerm2020enhanced}}\end{tabular}  & \begin{tabular}[c]{@{}l@{}} DL-based strategy to balance resource consumption between edge de- \\ vices and terminals  \end{tabular}& DL \\ \cline{3-5} 
			&   & {\cite{lim2021dynamic}} & \begin{tabular}[c]{@{}l@{}} A FL-based hierarchical framework of dynamic resource allocation f- \\ or terminals \end{tabular} & FL\\ \cline{2-5}
			
			& \multirow{3}{*}{\begin{tabular}[c]{@{}c@{}} Low latency \\ optimization\end{tabular}} & \begin{tabular}[c]{@{}c@{}}{\cite{chen2021matching}},  {\cite{zhu2020broadband}} \end{tabular}  & \begin{tabular}[c]{@{}l@{}} Optimization of EL network structure for latency reduction \end{tabular}& DL and FL \\ \cline{3-5} 
			&   & \begin{tabular}[c]{@{}c@{}}{{\cite{dinh2021federated}}, \cite{Skatchkovsky2019optimizing}} \\ \cite{prakash2021coded}  \end{tabular} & \begin{tabular}[c]{@{}l@{}}Acceleration of EL network convergence for latency reduction\end{tabular} & \begin{tabular}{@{}c@{}} DL and FL \end{tabular}         \\ \cline{2-5} 
			
			& \multirow{1}{*}{\begin{tabular}[c]{@{}c@{}}\\ Energy efficiency \\ optimization\\  \end{tabular}}  & {\cite{zeng2020energy}, \cite{yang2021energy}} & \begin{tabular}[c]{@{}l@{}} Efficient resource allocation and user selection to enable energy-effic-\\ient FL implementation\end{tabular}                                 & FL          \\ \cline{3-5} 
			&   & {\cite{vu2020cellfree}} & \begin{tabular}[c]{@{}l@{}}Joint power and rate control under imperfect channel information\end{tabular}                                                                                          & FL     \\ \cline{3-5} 
			&  & {\cite{yang2020energyefficient},\cite{irtija2022energy}} & \begin{tabular}[c]{@{}l@{}}A framework of DL-based edge processing for communication \\ power minimization\end{tabular}    & DL     \\  
			
			\hline
			
	\end{tabular}}
\end{table*}

To achieve differential privacy in distributed DL, a framework was proposed in \cite{du2020approximate} based on a new approximation mechanism while considering practical communication restrictions in the actual system, like bandwidth restriction.
On the other hand, to realize local differential privacy for DL, it was proposed in \cite{arachchige2019local} that all edge users add a randomization layer to convolutional neural networks (CNN).

Concerning FL, it is thought of a privacy-secure architecture \cite{konevcny2015federated} \cite{mcmahan2017communication}. A typical model update strategy in a federated network is 
\begin{equation}
	\bm{\theta}_{t+1} = \bm{\theta}_t - \eta_t\frac{1}{K}\sum_{k=1}^K{\bm g}_k(\bm{\theta}_t),
\end{equation}
where $\eta_t$ is the learning rate, subscript $t$ denotes the iteration index, ${\bm g}_k(\bm{\theta}_t)$ is the gradient computed at the $k$-th user. 
Because edge users only transported the gradient information of local model, i.e., $\mathbf{g}_k$ in (8), to the central server, FL was considered sufficient to protect the privacy of users.
However, it is recently found in \cite{zhu2019deep} that a neural network based on RL is able to break the privacy protection of FL. 
Therefore, for the FL framework, additional methods for privacy protection should be considered.
An ingenious way to add privacy protections in the FL framework was proposed in \cite{liu2020privacy}. 
In particular, the authors first proved that channel noise can be used to achieve differential privacy in FL.
In addition, another method to provide privacy protection was proposed in \cite{tabassum2021privacy} by incremental learning in network intrusion detection systems (NIDS).
The incremental learning reprocess the transmitted data, which led to a distribution of the input classifier data different from the original data, so as to protect the data privacy of users.  

\subsubsection{Security Design}
On top of the privacy protection, security of edge networks is another challenging but important factor that needs to be considered seriously. 
Due to the limited computational power, memory capacity, battery life, and network bandwidth of edge devices deployed on the Internet, the edge networks are facing endless threats or attacks. 

In particular, the study \cite{chaabouni2019network} analyzed the threats and challenges faced by Internet-of-Things (IoT), including impersonation attacks, distributed Denial of Service (DDoS) attacks, routing attacks, etc. 
Meanwhile, in \cite{chaabouni2019network}, the authors also introduced traditional defense mechanisms which protect the current IoT, including filter packets, adopting encryption, audit and log activities, etc. 
A DL-based method was proposed in \cite{chaabouni2019network} to protect network security in NIDS, including building free datasets for NIDS implementation, monitoring the network transit traffic using free and open-source network sniffers, and using open-source NIDS tools for detecting malicious events. 

On the other hand, the authors of \cite{ferrag2021federated} studied the performance of FL approaches for cyber security in IoT, such as detecting compromised IoT devices. 
Specifically, three different FL networks were considered with DNN, CNN, and recurrent neural network (RNN) architectures, which validates that FL framework is helpful for guaranteeing security in IoT. 

In addition to the above terrestrial communication scenarios, satellite communication will also be an important part of the future edge networks. 
However, due to limited computing resources in space, a popular practice, i.e., computation offloading (CO), in edge/fog computing is a potential solution, which alters the threat and risk profile of the system. 
In \cite{sthapit2021reinforcement}, a security-aware algorithm for CO was based on RL. 
In specific, the authors in \cite{sthapit2021reinforcement} formulated the security-aware CO problem as a multi-objective problem and designed a RL network to achieve secure satellite communication.

\subsection{Dual-functional Performance Metrics for Communication}

It has been stated that EL networks enable better communication to user terminals than the typical cloud data center method \cite{haojialu2020secure}. For the cloud data center method, the input data from user terminals is sent to a remote cloud data center and then the cloud data center feeds back execution results to user terminals. A large amount of data is transmitted back-and-forth between terminal devices and the cloud center over a wide-area network, which can result in high latency and excessive energy consumption. In EL, neural networks are trained on devices that are close to user terminals, which avoids sending data to the cloud center and therefore can reduce communication delay and support sophisticate ML algorithms with distributed computation offloading. Due to the diversity and heterogeneity of user terminal devices, researchers have devised numerous algorithms to optimize the communication metrics for EL networks, e.g., spectral efficiency, latency, and energy efficiency, to achieve efficient communication between user terminals in EL, details of which are summarized in Table III.

\subsubsection{Spectral Efficiency Optimization}

Similar to data transfer-oriented communication optimization, spectral efficiency is still a key performance metric in EL and edge computing optimization.  Recently, researchers have focused on spectral efficiency optimization by designing algorithms for edge computing and introducing AI to edge networks. Due to the heterogeneity of user terminals and the large amount of data processed at EL networks, it is challenging to improve spectral efficiency for EL networks on edge devices with limited computing power and storage capacity. By considering the heterogeneity of user terminals, a finite memory multi-state framework was proposed in \cite{park2019distributed} to share and reallocate limited communication resources in a self-adaption manner. The framework first identified periodic and critical messages,  then dynamically allocated communication resources based on the number of critical messages. To overcome the challenge of transmitting a large amount of data to the edge network, a three-layer network was proposed in \cite{wang2019joint} to jointly utilize communication resources in cloud centers, access point, and edge devices. 


With the development of AI, DL and FL techniques have been introduced to edge networks to optimize spectral efficiency. In \cite{chen2018label}, a DL-based control algorithm based on label-less learning was proposed to minimize the amount of data communication by using  limited computing and spectrum. An enhanced online Q-learning network was proposed in \cite{alqerm2020enhanced} to optimize the spectral efficiency and retain the fairness of  resource allocation simultaneously. This Q-learning network first received context information from user terminals, including priority, latency information, and server load. Then it allocated resource to edge devices by exploiting the above context information. On the other hand, by exploiting FL, a hierarchical game framework was proposed in \cite{lim2021dynamic} to study the dynamics of edge association and spectral efficiency in an edge network.

\subsubsection{Low Latency Optimization}

Another important metric of communication for EL is the end-to-end latency, especially for B5G networks. Note that the definition of latency in EL networks is different from traditional communication latency. Edge communication latency is the total time from the generation of demand by user terminals to the completion of calculation by edge devices. Researchers have paid attention to reduce the communication latency by optimizing the EL network structure and elevating the EL network convergence rate. For optimizing the EL network, a large-scale matching algorithm was proposed in \cite{chen2021matching} to find the optimal low latency assignment. Besides, a multi-access network for EL was proposed in \cite{zhu2020broadband} to reduce the communication latency. The multi-access network balanced receive SNR, truncation ratio, and a fraction of exploited data metrics to optimize the communication latency. In \cite{dinh2021federated}, an FL algorithm was enhanced to handle heterogeneous UE data and reduce the communication latency. Also, for elevating the EL network convergence rate, given the transmission overhead and transmission efficiency of each data packet, an optimal solution was proposed in \cite{Skatchkovsky2019optimizing} to accelerate network convergence. To address the issue of slow  convergence, a novel coded computing framework was proposed in \cite{prakash2021coded} to mitigate stragglers and speed up the training procedure by injecting structured coding redundancy into FL.

\subsubsection{Energy Efficiency Optimization}
As a counterpart of spectral efficiency, energy efficiency considers the average energy consumption of a communication system in order to transmit a unit of information. In general, EL approaches, e.g., FL, facilitates the training of models and enables distributed data collection \cite{konevcny2016federated}. However, due to the growing number of transmission nodes and unstable wireless channels with limited bandwidth, implementing distributed algorithms usually costs large energy consumption, which does not meet the demand on green communications and efficient computing. 

In order to address the challenge imposed by the need of energy-efficient FL, various approaches have been developed in recent years. In \cite{chen2020joint}, by considering the connection between the FL and resource allocation, a universal and flexible framework was proposed for enabling a practical implementation of FL models. To reduce the energy computation of the training of FL, a resource allocation algorithm was proposed in \cite{yang2020federated} to balance the training time of FL models and the energy consumption of UEs. In \cite{zeng2020energy}, an optimal bandwidth allocation policy with a closed-form scheduling priority function was presented to save the energy consumption of UEs without learning performance loss. To save the total energy consumption of a system under a latency constraint, the study \cite{yang2021energy} proposed a low-complexity energy-efficient algorithm, for solving classical resource optimization problems, e.g., bandwidth allocation. 

In an MEC-based wireless network, deploying DL models for FL, i.e., DNNs, and executing inference tasks are also very challenging for saving energy \cite{xu2018scaling}. To facilitate energy efficient transmission in an MEC-based wireless network, a framework of edge processing was proposed in \cite{yang2020energyefficient}, where DL inference tasks can be effectively executed at the edge computing nodes. In particular, the minimization of the sum of the computation in edge nodes and the power consumption for signal transmission is considered. A statistical learning-based robust optimization method was proposed for solving this minimization problem.

%% file: sections/section3-B.tex
\subsection{Communication Optimization for Edge Learning}

In EL networks, it is of great importance to properly manage limited wireless resources for implementing ML algorithms. More specifically, for FL, optimizations to the wireless resources, e.g., transmit power and frequency spectrum, of the edge devices can bring tremendous improvements to the entire system, e.g., achieving lower energy assumption \cite{yang2021energy,Yao2021Enhancing}, and accelerated convergence speed \cite{Chen2021Convergence}.
On the other hand, MEC, as an effective framework with the ability of distributed computing, can offload learning tasks and computing resources for EL. Thus, the strategy of task and computing resource allocation plays an important role to realize energy-efficient and low-latency MEC \cite{Xia2021Online,mao2017a}.
In this subsection, we provide a comprehensive overview of common optimization methods for communication in FL and MEC networks, respectively.

\subsubsection{Optimization of FL Networks}
In an edge FL network, jointly optimizing the decisions of multiple devices can improve the learning efficiency of the collaborative system.
In general, the involved optimization problem can be modeled as follows:
\begin{align}\label{prob4FL}
\mathop \text{minimize} \limits_{ \mathbf x} \quad
& F(\mathbf x) \triangleq \sum_{k=1}^K f_k(\mathbf x, \mathbf \xi)  \\
\text{subject to}\quad
& \mathbf x_k \in \mathcal X_k , \ k = 1, \cdots, K, \tag{\ref{prob4FL}{a}} \label{prob4FL:c1} \\
& h_m(\mathbf x, \mathbf \xi) \leq 0 , \ m = 1, \cdots, M,  \tag{\ref{prob4FL}{b}} \label{prob4FL:c2}
\end{align}
where $\mathbf x \triangleq \{\mathbf x_k \}_{k=1}^K$ denotes the set of optimization variables of all involved devices and $\mathbf x_k \in \mathcal X_k$ corresponds to the decision variable judged at the $k$th device with $\mathcal X_k$ denoting the corresponding local set. Here $\mathbf \xi$ represents the set of problem parameters relying on the network environments, e.g., CSI. The function $F(\mathbf x)$ is the design objective of the edge network consisting of a series of local functions of all devices, $f_k(\mathbf x, \mathbf \xi), \ k = 1, \cdots, K $. The involved constraints can be categorized as the local individual constraints at each device in (\ref{prob4FL:c1}), and the network-level constraints in (\ref{prob4FL:c2}). The former is locally associated with each device and the latter is used for guaranteeing a cooperative design of the entire network.

Some of the problems in (\ref{prob4FL}) can be addressed using convex optimization techniques. Concretely, by splitting the variables into several groups and employing the alternating optimization framework, the original complicated problem can usually be transformed into multiple convex subproblems and solved in an iterative manner.
For each of these convex subproblems, by further analyzing the properties of the objective function, e.g., the monotonicity, or focusing on manipulating the Karush-Kuhn-Tucker (KKT) conditions, a closed from or semi-closed form solution can be obtained, thus yielding a relatively low computational complexity.

For example, the authors in \cite{zeng2020energy} investigated
the energy consumption minimization problem for implementing FL over wireless channels via iteratively optimizing the bandwidth allocation and user scheduling. With a given set of active devices, the subproblem of bandwidth allocation can be formulated considering the following constraints
\begin{align}
\sum_{k=1}^K \gamma_k=1,
\quad \ 0\leq \gamma_k \leq 1, \ k=1,\cdots,K,
\end{align}
where $\gamma_k$ denotes the ratio of bandwidth allocation for device $k$, and $0\leq \gamma_k \leq 1, \ k=1,\cdots,K$ are individual constraints for each device while $\sum_{k=1}^K \gamma_k=1$ is a system-level constraint.
By directly solving the KKT conditions, a group of closed form solutions to $ \{\gamma_k\}_{k=1}^K$ of the bandwidth allocation subproblem can be obtained.
In contrast with the classical design for rate maximization, the optimized results of \cite{zeng2020energy} indicate that
more bandwidths should be allocated to those scheduled devices with weaker channels and worse computation capacities, since they are the bottlenecks for synchronized model updates in an edge FL system.
Similarly, a joint computation and transmission problem was studied in \cite{yang2021energy} for edge FL networks under a latency constraint. With the assistance of the proposed iterative algorithm, the authors derived closed form solutions for assigning time and bandwidth, power control, computation frequency, and even learning accuracy of each device at each iteration.
In \cite{Yao2021Enhancing}, the authors studied the trade-off between energy consumption and learning time of FL in fog-aided IoT networks. They proposed an alternating optimization algorithm to optimize CPU frequency and wireless transmission power, where a closed form solution and a semi-closed form solution were obtained, respectively.
On the other hand, the work in \cite{Amiri2020Federated} considered the implementation of distributed stochastic gradient descent (SGD) for FL and an efficient power allocation scheme was given for aligning the received gradient vectors at the parameter server.

The above convex optimization-based methods provide efficient solutions of the optimization in FL networks. However, it is applicable only when the considered problem is relatively simple. Unfortunately, for most cases, there exists complicated coupling among variables and strong non-convexity in problem (\ref{prob4FL}) such that conventional convex optimization approaches fail.
To tackle this difficulty, a natural and direct method is to replace the non-convex functions by convex approximations.
For example, the authors of \cite{vu2020cellfree} proposed a scheme for implementing FL in massive MIMO networks, where each iteration of the FL framework is accomplished during a large-scale coherence time. Then, an FL training time minimization problem using this proposed scheme was exemplified as a case study.
The local accuracy, transmit power, data rate, and computational frequency were jointly optimized based on a successive convex approximation (SCA) method, by solving a sequence of convex problems.
In \cite{Liu2021Reconfigurable}, the authors considered a reconfigurable intelligent surface (RIS)-assisted over-the-air FL network. The non-convex problem of the joint design of receiver beamforming and RIS phase shifts was solved by exploiting the technique of semidefinite relaxation (SDR) \cite{Luo2010Semidefinite}.
Moreover, for non-convex unit-modulus phase shift constraints in RIS-aided systems, the majorization-minimization (MM) framework \cite{Sun2017MM} and the Riemannian manifold optimization \cite{ma2012manifold} are also commonly used.


Besides the above resource allocation elements with continuous values, device selection is usually necessary for the BS to extract appropriate devices so as to execute the FL algorithm, since the bandwidth for multiple users uplink transmission is limited. This leads to a mixed integer optimization problem. Mathematically, the device selection is taken into account by multiplying a series of integer factors, written as
\begin{align}\label{cons:MixInte}
a_k \in \{0,1\},\  k=1,\cdots,K,
\end{align}
to weight each device, where $a_k = 1$ indicates that user $k$ participates the FL algorithm and otherwise $a_k = 0$.
For handling this kind of mixed integer optimizations, the works \cite{Wang2022Federated,yang2020federated} transformed the device selection problem to a reformulated sparse and low-rank optimization problem.
Problems of sparse optimization and low-rank optimization occur frequently in ML and signal processing \cite{Shi2014Group,Davenport2016An,Shi2018Generalized,yonina2015Sampling}, whose difficulties mainly lie in the minimization of the nonconvex sparse function, $\| \mathbf x\|_0$, and the low-rank constraint of a positive semidefinite matrix, $\text{rank}(\mathbf M ) = 1$, where $\text{rank}(\cdot)$ returns the rank of the input matrix $\mathbf{M}$.
In previous works, the non-convex sparsification is often approximated by the convex $\ell_1$-norm or the smoothed $\ell_p$-norm minimization \cite{Shi2016Smoothed} and the technique of SDR is widely used to handle the rank-one constraints.
Different from these methods, the works \cite{Wang2022Federated,yang2020federated} developed a unified difference-of-convex-functions (DC) programming based approach to deal with sparse and low-rank optimizations in FL networks with global convergence guarantees, which yields considerable performance improvements.
Moreover, the authors of \cite{Chen2021Convergence} proposed a probabilistic device selection scheme aiming at choosing the devices, whose local learning models have larger effects on the global model, with higher probabilities.
In \cite{Shi2021Joint} and \cite{Liu2021Reconfigurable}, a greedy device scheduling algorithm and a Gibbs sampling based device selection method were devised, respectively.

\subsubsection{Optimization of Edge Computing/Caching}
Another type of edge network is for edge computing/caching.
The optimizations of resource management in MEC edge networks have been discussed in \cite{mao2017a}.
In particular, stochastic optimization is an important focus since the CSI acquisition in MEC is inevitably imperfect owing to channel estimation error \cite{jindal2010unified}, limited feedback \cite{love2008overview}, uncertainty \cite{shi2014optimal}, and delays \cite{maddah2012completely}.
With the stochastic CSI and unknown link conditions, stochastic optimization can adopt online decisions to achieve optimal solutions in MEC networks.
For example, a Lyapunov stochastic optimization based algorithm was proposed in \cite{han2020joint} to jointly optimize the transmission rate and computation rate for minimizing power consumption. Then, based on game-theoretic and perturbed Lyapunov optimization theory, the authors in \cite{Xia2021Online} jointly optimized task offloading, computing resource allocation, and battery energy management in a distributed energy harvesting-enabled MEC system.
Moreover, alternating direction method of multipliers (ADMM) is another approach to distributed stochastic optimization in EL integrated IoE. A coding-based stochastic ADMM algorithm was proposed in \cite{chen2021coded} to optimize the communication efficiency and straggler nodes in coded edge computing networks.
%
%
%
%
%
%
\subsection{Edge Learning for Communication Optimization}
Techniques of EL benefit from the optimization of wireless resource allocation. On the other hand, learning is useful for solving complicated optimization problems in edge communication networks. ML-driven approaches can overcome the drawbacks of conventional optimization methods, such as numerous iterations and high computational complexity. It has been applied to power 
allocation \cite{sun2018learning}, precoding design \cite{lee2020deep}, and other end-to-end designs in communication systems. The EL techniques including DL, FL, and RL are regarded as promising approaches for solving resource allocation problems in edge computing, IoE, and other edge networks. We introduce learning-driven edge network optimization with perfect CSI and statical CSI, respectively.  

\subsubsection{EL Techniques with Perfect CSI}
Learning-based distributed optimization is an essential technique in communication systems, especially in edge networks like MEC. To address the lack of latency-energy balance, time efficiency, and robustness in MEC, a number of studies have focused on ML-based computation offloading algorithms \cite{huang2018learning,zhu2019novel}. 
In general, the optimization problem in MEC networks can be formulated as
\begin{align}\label{probML}
\mathop \text{minimize} \limits_ {\bf{X}} \quad
& U\left( {\bf{X}} \right) \buildrel \Delta \over = {\lambda _t}T\left( {\bf{X}} \right) + {\lambda _e}E\left( {\bf{X}} \right)   \\
\text{subject to} \quad
& {x_{ij}} \in {\cal X},\quad \forall i \in {\cal I},  \forall j \in {\cal J}, \tag{\ref{probML}{a}} \label{probML:c1} \\
& g\left( {{{\bf{x}}_i},{\tau _i}} \right) = {M_i}, \quad \forall i \in {\cal I}, \tag{\ref{probML}{b}} \label{probML:c2}  \\
& h\left( {{{\bf{x}}_j},{\tau _j}} \right) \le {N_j}, \quad \forall j \in {\cal J},
\tag{\ref{probML}{c}} \label{probML:c3}
\end{align}
where ${\bf{X}} \buildrel \Delta \over = \left\{ {{x_{ij}}} \right\}$ denotes the variables of MEC task offloading indicators and ${\cal I}$ and ${\cal J}$ are respectively the set of mobile devices (MD) and the set of computation access points (CAP).
Each element ${x_{ij}}$ denotes the variable with respect to MD $i$ and computing CAP $j$. The utility function $ U\left( {\bf{X}} \right)$ can be formulated as a weighted objective function representing the balance between $ T\left( {\bf{X}} \right)$ and $ E\left( {\bf{X}} \right)$, which respectively denote the two key metrics in MEC as edge computing latency and power consumption.
The optimization variable ${x_{ij}}$ is characterized by the policy space ${\cal X}$, e.g., integer constraints. Equations (\ref{probML:c2}) and (\ref{probML:c3}) denote the constraint of MDs and CAPs, respectively, where ${\tau}$ denotes the system parameter set in the MEC system.

Rather than resorting to convex optimization tools for solving (\ref{probML}), ML-based methods can be utilized for getting near-optimal solutions to (\ref{probML}) with high probability and low complexity \cite{huang2018learning,zhu2019novel}. By regarding the solution of $\bf{X}$ as a random variable following a specific probability mass function with respect to the given system parameters, it is equivalent to solve (\ref{probML}) by learning the probability mass function of optimal $\bf{X}$ in a data-driven way using ML algorithms. Denote by $q\left( {\bf{X}} \right)$ as the probability mass function of the optimal solution of $\mathbf{X}$, and let $p\left( {\bf{X}} \right)$ be the probability distribution function of $\bf{X}$ that is to be learned. The problem of (\ref{probML}) is equivalent to finding $p$ in the following problem as
\begin{align}\label{probCE}
\mathop \text{minimize} \limits_{ p} \quad
& \sum {q\left( {\bf{X}} \right)\ln q\left( {\bf{X}} \right)}  - \sum {q\left( {\bf{X}} \right)\ln p\left( {\bf{X}} \right)}.
\end{align}
The objective function is defined by the cross-entropy (CE) \cite{de2005tutorial} measuring the difference between ${p\left( {\bf{X}} \right)}$ and ${q\left( {\bf{X}} \right)}$. The optimization problem (\ref{probCE}) is a probability approximation problem that is readily solved by ML methods with offline training.

Another form of learning-based MEC design exploits DL methods to learn the mapping function of an arbitrary problem in edge networks. A general framework of deep learning integrated optimization was proposed to solve non-convex problems in wireless resource management \cite{lee2019deep}. The framework was implemented in a distributed paradigm, where multiple DNNs were utilized to work as compute nodes and exchange information via backhaul. Specifically, a distributed DL-based algorithm was proposed in \cite{wu2020collaborate} to optimize task offloading strategies in a heterogeneous network of cloud and edge computing.
However, DNNs in the above studies are exploited to learn the solution variables directly from the training dataset. To address the lack of generalization capability of conventional DL methods, a DNN-based offloading assignment method was proposed in \cite{qian2022learning} to learn the pruning strategy as a part of the entire algorithm instead of learning the entire offloading strategy directly. It achieves low complexity, sufficient robustness and near-optimal efficiency performance in the tested multi-user MEC network. 
In addition, learning-driven approaches without DNN have also been applied to edge networks. A low-rank learning-based algorithm was proposed in \cite{hu2019learning} to predict task execution time with the knowledge of a small sampled dataset, then a task offloading algorithm based on this predicted task execution time was proposed to improve the success rate of task offloading and reduce latency in the edge computing network. To improve the fairness of users, a multi-agent imitation learning scheme was utilized in \cite{wang2020multi2} to optimize the computation offloading strategy in a fully decentralized pervasive edge computing network. 

The techniques of RL have also been shown promising in solving optimization problems in edge networks, owing to its adaptive capacity in dynamic environments. Instead of solving a single problem of computation offloading, many studies intended to jointly optimize offloading, caching, resource allocation, security, and other issues via RL-based methods. Edge caching and resource allocation were jointly considered in \cite{zhang2021joint}, where a DRL-based algorithm was proposed to design caching strategies. In addition, a Bayesian DL-based method combined with DQL was proposed in \cite{asheralieva2019distributed} to jointly optimize the pricing and resource management in a blockchain integrated edge computing networks. 

As an alternative, distributed FL has also been  extensively employed in edge network optimization. A distributed multi-agent deep deterministic policy gradient algorithm realized by FL was proposed in \cite{kwon2020multiagent} to jointly decide resource allocation and cell association in ocean IoT environment. The learning-based joint optimization in \cite{kwon2020multiagent} was modeled in a single timescale. However, the 
various delay sensitivity of caching, computation offloading, and resource allocation can be described by different timescales.
An FL-based approach was proposed in \cite{yu2020deep}  to jointly optimize resource allocation, offloading, and caching to reduce latency and save resource consumption in MEC. In \cite{yu2020deep}, caching was regarded as delay insensitive and managed in slow timescale, while the other issues were managed in fast timescale. The FL-based training helps ensure the privacy of information in edge devices.

\subsubsection{EL Techniques with Statical CSI}
Communication optimization with only statical CSI is a crucial challenge in distributed networks due to imperfect CSI acquisition and dynamically changing network topology. Learning techniques can overcome the lack of adaptability to stochastic wireless environments in conventional optimization methods. 
Conventional learning-based methods are designed with the assumption of a single distributed dataset, which is hardly scalable in practical scenarios. Thus, it is essential to design learning-based models that fit the dynamics and uncertainty of CSI. 
In particular, a model-free DQL framework was proposed to apply the dynamic power allocation strategy sum rate maximization with scheduling in a mobile Ad-hoc network \cite{nasir2019multi}. Specifically, this unsupervised DQL-based method with novel designs of state and reward was shown to obtain near-optimal performance.  

Learning-based stochastic optimization is also considered in edge computing networks. For edge inference with finite samples of random channel coefficients, a statistical learning-based approach was exploited to approximate the robust optimization of cooperative transmission \cite{yang2020energyefficient}. Since the optimization problem could not be expressed in a closed form due to the joint chance constraints in MEC, robust optimization approximation and statistical learning-based approaches can provide a robust and energy-efficient solution by learning the parameters from a finite dataset. 
Also, FL is promising for solving distributed stochastic optimization problems in edge networks. An FL-based joint scheduling and resource allocation algorithm under imperfect CSI was proposed in \cite{wadu2021joint}. The FL-based stochastic optimization algorithm predicts the unexplored CSI via Gaussian process regression and dual-plus-penalty. This improved the accuracy of FL and the stochastic algorithm was shown to be robust against various CSI distributions. 

%% file: sections/section3-C.tex
\subsection{Convergence, Complexity, and Signaling Overhead}

\begin{table*}[ht!]
\caption{Summary of Convergence Analysis}
\label{convergence_tab}
\centering
\begin{tabular}{|c|m{2.8cm}<{\centering}|m{9.5cm}|m{2.7cm}|} \hline
\textbf{Reference}& \textbf{Loss function} &  \multicolumn{1}{c|}{\textbf{Factors of the framework}} & \multicolumn{1}{c|}{\textbf{Convergence rate}}\\ \hline
\cite{zhu2020one}& Convex and nonconvex&  One-bit gradient quantization, AirComp, fading channels, perfect/imperfect CSI&$\mathcal{O}(1/\sqrt{T})$\\ \hline
\cite{yang2019scheduling} & Strongly convex &  Transmission scheduling policy, features of wireless channels, inter-cell interference &$\mathcal{O}\left(\log(\frac{n}{\epsilon})\right)$ \\ \hline
\cite{chen2020fedcluster} & Nonconvex & Grouping devices into $M$ clusters, cluster-cycling, device-level data heterogeneity & $\mathcal{O}(1/\sqrt{MNT})$\\ \hline
\cite{liu2021joint} & Nonconvex & Model pruning, device selection, wireless resource allocation & $ Slower than \mathcal{O}(1/\sqrt{T})$ \\ \hline
\cite{fan2022bev} & Convex and nonconvex& AirComp, robust transmission policy against Byzantine attacks & $\mathcal{O}(\Omega / ( \omega^2 \sqrt{T}) )$\\ \hline 
\cite{liu2021resource} & Convex & Distributed approximate Newton-type algorithm, heterogeneous and non-i.i.d. data & $\mathcal{O}\left(\log(\frac{1}{\epsilon})\right)$ \\ \hline
\cite{liu2021resource} & Nonconvex & Distributed approximate Newton-type algorithm, heterogeneous and non-i.i.d. data &  $\mathcal{O}(1/\sqrt{T})$ \\ \hline
\cite{chen2022distributed} & Strongly convex &  Sparsification and error correction, sparsified gradient difference transmission&$\mathcal{O}\left(\log(\frac{1}{\epsilon})\right)$ \\ \hline
\end{tabular}
\end{table*}

\subsubsection{Convergence}
Convergence is one of the key factors determining the accuracy, rate, and overhead of EL. Through theoretical convergence analysis, the impact of wireless factors on convergence can be specified, which guides the optimization design for edge networks. We discuss state-of-the-art convergence analysis of optimization schemes under realistic constraints, including heterogeneity of data and devices, dynamic wireless environments, and limited communication resources.\par

To begin with, we focus on a decentralized network with a central coordinator, a.k.a. a parameter server as depicted in Fig. 2. FL is one of the typical representatives of this network of EL and its convergence rate achieves $\mathcal{O}(1/\sqrt{NT})$ for nonconvex loss functions \cite{yu2019parallel} and $\mathcal{O}\left(\log(\frac{1}{\epsilon})\right)$ for strongly convex ones \cite{ma2017distributed}, where $N$ and $T$ represent the number of devices and the number of iterations, respectively, and $\epsilon>0$ is the required accuracy. Considering the impact of unreliable communication, there may exist a gap between the convergence of the model in practice and the optimal one in theory. Given the presence of uplink transmission errors, the study in \cite{chen2020joint} derived the expected convergence rate and revealed that the packet errors lead to a gap between the globally optimal model. In \cite{wang2019adaptive}, reducing the frequency of global aggregations was considered and the gap in term of convergence rate under this scheme was characterized. Over-the-air computation (AirComp) is another effective way for facilitating FL with communication constraints. For example, in \cite{fan2021joint}, the expected convergence rate for both convex and non-convex cases was derived, which accurately revealed the influence of AirComp on convergence. In addition, limited communication resources may also have an impact on the convergence rate of the model. Therefore, methods to reduce communication overheads and speed up convergence have been widely discussed. For FL with data heterogeneity, fast convergence could be achieved by exploiting nonuniform aggregation of the gradients from different devices \cite{nguyen2020fast}. Similarly, a node selection method in \cite{wu2021node} was also shown to have a faster convergence rate in the face of non-i.i.d. data. In \cite{liu2021hierarchical}, a hierarchical FL system with less aggregations and quantization was proposed and the convergence rate of $\mathcal{O}(1/\sqrt{T})$ was derived for non-convex loss functions. The authors in \cite{sun2020toward} exploited the gradient sparsification combined with gradient correction and batch normalization (BN) update with local gradients to reduce communication overheads and accelerate the convergence. \par

Further considering a fully distributed network with no central coordinator, where the devices can only communicate with finite neighbours. Under vulnerable communications, the authors in \cite{ye2021decentralized} adopted the user datagram protocol (UDP) for more efficient transmission and developed a robust algorithm with asymptotic convergence rate of $\mathcal{O}(1/\sqrt{NT})+\mathcal{O}(N/T)$. Furthermore, in \cite{elgabli2020gadmm}, an ADMM-based communication-efficient framework was proposed which is able to guarantee optimality of convergence under a convex loss function. Different from the supervised learning methods discussed above, MARL is also a typical distributed EL technique, which can achieve optimal decision by interacting with the dynamic environments \cite{zhang2021multi,hu2021distributed}. Unfortunately, to the best of our knowledge, a rigorous analysis of the convergence of MARL has not yet been reported in literature.\par

Except for the above representative schemes, we summarize the rest of the latest research results about convergence in Table~\ref{convergence_tab}.

\subsubsection{Complexity}
Optimization for EL grants better performance via rational allocation of resources at the expense of computational complexity in practice. Excessive computational complexity introduces larger latency, thereby reducing the performance gain brought by the proposed optimization. \par

To begin with, we consider methods based on convex optimization tools, which are widely used for continuous variable optimization under perfect CSI. Based on an iterative mechanism, low-complexity algorithms are available by deriving closed-form solutions of convex subproblems \cite{yang2020energy2,wang2021asynchronous}. The computational complexity mainly depends on the accuracy requirement of convergence as well as the calculations of the closed-form solutions. However, most practical problems admit intractable non-convex forms. When it is difficult to obtain closed-form solutions of the subproblems, it may be solved by means of, e.g., the interior point method, with polynomial computational complexity. For example, in \cite{yang2020energy2}, due to the introduction of semi-definite programming (SDP), the complexity of solving subproblems in each iteration is $\mathcal{O}(N^6)$, where $N$ is the the dimension of the problem. In addition, to handle discrete variables, mixed integer optimization is applied with its complexity dominated by the size of the solution space. In \cite{fan2021joint}, the complexity of the problem was $\mathcal{O}(U)$ and the size of the search space, $U$, was further reduced without loss of optimality.\par

In order to reduce complexity, learning-based methods can be computationally efficient for some cases. For example, a graph neural network (GNN) was applied to assist radio resource management in \cite{shen2020graph} and its complexiy is $\mathcal{O}(LD)$, where $L$ is the number of layers of the GNN and $D$ is the maximal degrees of the graph. Compared with the conventional weighed minimum mean-square error (WMMSE) method in \cite{shi2011iteratively}, the GNN-based scheme has a significant complexity reduction. In \cite{qian2022learning}, DL was applied to optimize resource assignment in a multiuser MEC system and the complexity was reduced by 80\% compared to the conventional branch-and-bound approach.

\begin{figure*}[!htp]
\centering
\includegraphics[width=7.0in]{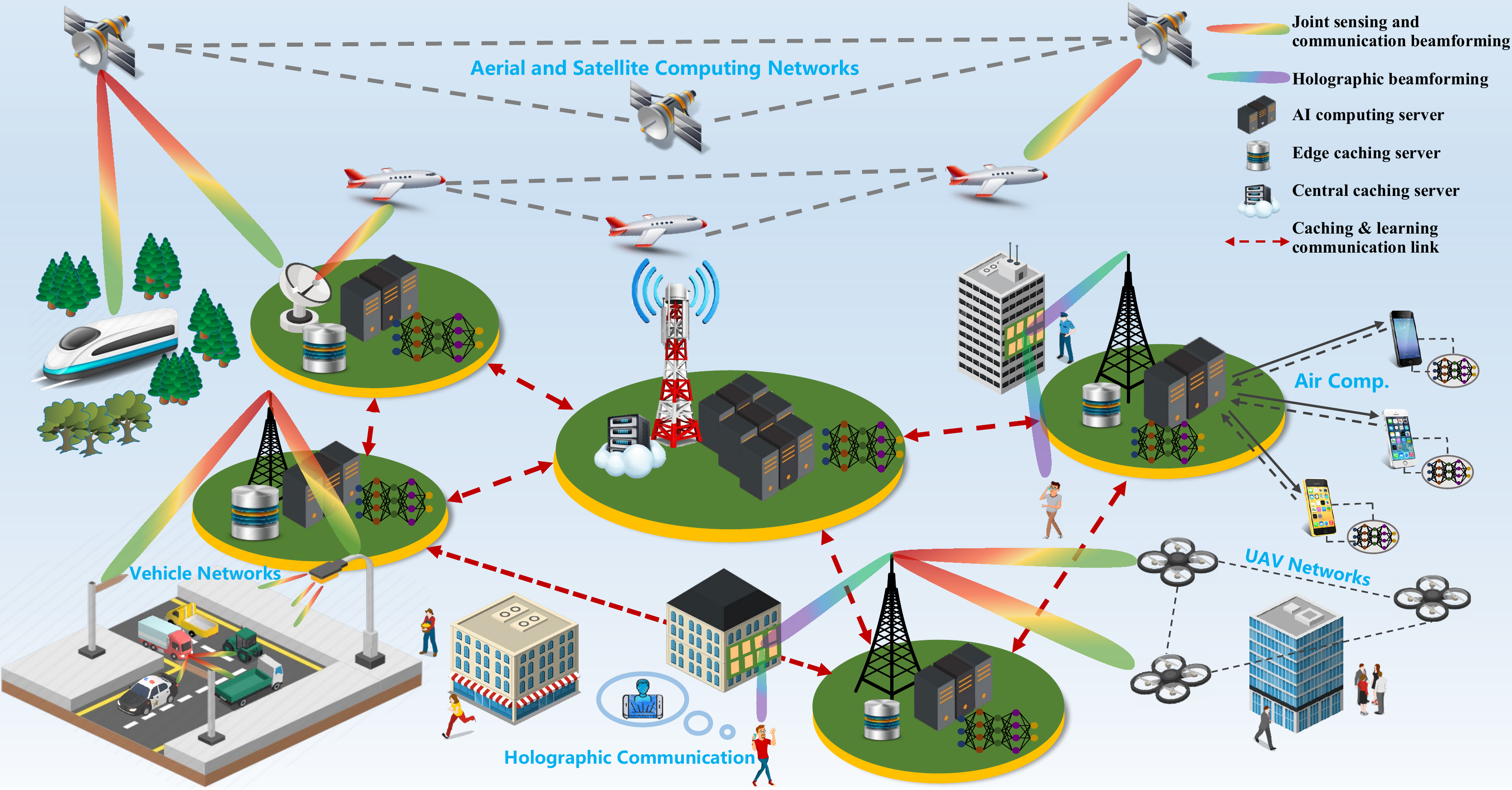}
\caption{Some typical application scenarios of EL.} \label{fig4first}
\end{figure*}
 
\subsubsection{Signaling Overhead}
Due to limited communication resource, signaling overhead in implementation of EL is also an essential factor worth considering. A large number of signaling interactions lead to huge communication overheads and communication latency. They also greatly hinder the rate of model training and convergence.\par 

In a typical FL framework, the signaling overhead can be evaluated as 
\begin{align}
\mathcal{T}_{{\mathrm{FL}}}=2TPK,
\end{align}
where $T$, $P$, and $K$ represent the number of communication rounds, the number of model parameters, and the number of edge devices, respectively. It is not difficult to find that the signaling overhead is extensive with either a large model or a slow convergence rate. Many methods have been proposed to reduce the requirements on communication resource. To reduce the number of model parameters, $P$,  sparsification \cite{sun2020toward, luo2021novel} and quantization \cite{haddadpour2021federated, magnusson2020maintaining} have been widely used. As exemplified in \cite{luo2021novel}, sparsification is a largely effective compression method, which achieves a compression rate up to 1/600, while a quantization method can only achieve 1/32. For the number of communication rounds, the lazy aggregation scheme was verified to be effective in \cite{wang2019adaptive}. Furthermore, speeding up the convergence is also a useful method, such as those in \cite{liu2020accelerating,liu2021resource}. \par

For SL, all the values of gradients, tensor outputs from intermediate layers, and the labels need to be transmitted over wireless links. It leads to the total overhead as \cite{singh2019detailed}
\begin{align}
\mathcal{T}_{{\mathrm{SL}}}=2Tpq,
\end{align}
where $p>0$ and $q>0$ denote the total dataset size and the size of the smashed layer, respectively. Compared with FL, SL usually enjoys a faster convergence rate \cite{vepakomma2018split} and it is more suitable for situations with a massive number of edge devices. In this sense, SL may be a more communication-efficient architecture than FL.\par

For MARL, all values of reward, action, state, and the model parameters should be shared among different agents. For various MARL schemes, the information to be shared can be quite different, which results in different signaling overheads. 


For the procedure of DI, especially in latency-sensitive applications, signaling overhead can not be ignored. Similar to SL, in a device-server co-inference framework, a large DNN is divided into two parts, which are respectively deployed on the device and the server. Based on \cite{shao2020communication}, the choice of the split point determines the signaling overhead and also involves the trade-off between the communication overhead and the computational cost at the device. Recently, some possible methods to alleviate excessive signaling overhead have also been studied from the perspective of compression coding and reducing the scale of the model, such as joint source-channel coding (JSCC) \cite{jankowski2020deep} and pruning \cite{shao2020communication, jankowski2020joint}.

%% file: sections/section4-A.tex

\section{B5G Wireless Applications with Edge Learning}
In this section, we introduce multiple emerging application scenarios of EL, e.g., vehicle networks, UAV networks, satellite networks, over-the-air computation, and holographic communication, as depicted in Fig. \ref{fig4first}, where edge nodes implement the function of communication, sensing, caching, computing by utilizing ML algorithms.

\subsection{Distributed Goal-oriented Semantic Communication}

The entropy and capacity defined by Shannon\cite{shannon1948mathematical} characterizes the maximum achievable rate bound for the communication whose goal is to exactly recover the transmitted information at the receiver.
In most applications at IoE devices in B5G networks, however, the goal of communication is to make a correct inference or acquire a computation result from the received data at the server.
For these computing tasks, it has been shown in \cite{weng2021semantic,jankowski2020wireless,xie2021task} that the transmission rate can be further reduced below the entropy of the source and joint source-channel coding can perform better than a conventional separate coding design in terms of computing accuracy and speed.

\begin{figure*}[t]
\centering
\includegraphics[width=7.0in]{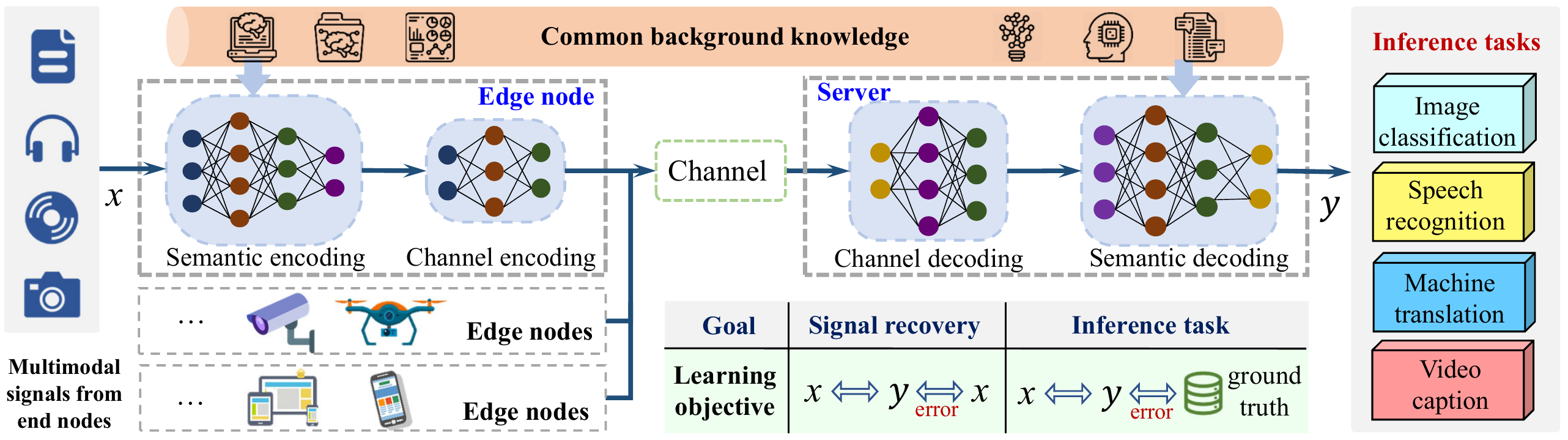}
\caption{Illustration of goal-oriented semantic communication system.} \label{fig4-0}
\vspace{-0.5em}
\end{figure*}

To provide the source entropy related to various task goals, {\em a first information theoretic model for the goal-oriented semantic communication} is proposed, where the source is $X$ and the desired information at the receiver is $Y$, as shown in Fig. \ref{fig4-0}.
In conventional communication for data recovery, it corresponds to the goal of realizing $Y = X$. Accordingly, the minimum required transmission rate is known to be ${\rm H}(X)$, that is the entropy of the source.
For a general task goal, $Y$ can be a decision or a prediction result that obeys some joint probability distribution ${\rm P}_{X,Y}(x,y)$, where $x$ and $y$ are realizations of $X$ and $Y$, respectively. Then, the goal-oriented communication can transmit at the minimum rate characterized by the following problem:
\begin{align}\label{goaloriented}
\mathop \text{minimize} \limits_ {g(\cdot)} \quad
& {\rm H}(\Tilde{X})   \\
\text{subject to} \quad & \Tilde{X} = g(X), \tag{\ref{goaloriented}{a}} \\
& {\rm I}(\Tilde{X};Y) \ge {\rm I}(X;Y), \tag{\ref{goaloriented}{b}}
\end{align}
where $g(\cdot)$ is a deterministic mapping function from $X$ to the compressed information $\Tilde{X}$, and ${\rm I}(\cdot;\cdot)$ is the mutual information of the two random variables, which is defined by the joint probability distribution ${\rm P}_{X,Y}(x,y)$ and the marginal distributions $P_X(x)$ and $P_Y(y)$.
In the above formulation, we seek a deterministic mapping, $g$, of the source $X$. It is used at the receiver to obtain a corresponding $Y$ from $\Tilde{X}$, while the accuracy should be, at least, the same as the best inference of $Y$ directly from $X$.
It is easy to verify that
\begin{align}  \label{transm_rate}
    {\rm H}(\Tilde{X}) \  \le & \  {\rm H}(g(X),X)   \nonumber    \\
                           =  & \  {\rm H}(X) + {\rm H}(g(X)|X)   \nonumber   \\
                           =  & \  {\rm H}(X),
\end{align}
where the last equality holds because $g$ is a deterministic function of $X$. This relationship in \eqref{transm_rate} proves that the minimum transmission rate of this goal-oriented communication design is theoretically no larger than the source entropy in the conventional design. In addition, it is expected that the inequality constraint in \eqref{goaloriented} achieves equality at the optimum, and this optimization problem can be further extended for modelling multi-node semantic communications in distributed edge networks.

In order to solve \eqref{goaloriented}, it is necessary to know the task goal in terms of ${\rm P}_{X,Y}(x,y)$ with explicit expressions. However, for most advanced applications like image retrieval and natural language processing, ${\rm P}_{X,Y}(x,y)$ can be hardly acquired with an explicit expression while it is usually learned implicitly by using DL methods. Without an explicit expression of ${\rm P}_{X,Y}(x,y)$ for these tasks, it is intractable to solve the problem in \eqref{goaloriented} by using conventional convex optimization tools. Therefore, recent works seek powerful DL techniques to obtain $g$ implicitly.
For instance in \cite{weng2021semantic}, a DL-based semantic communication system was proposed for speech transmission, where the mean-squared error (MSE) was used as the learning goal.
In \cite{jankowski2020wireless}, a retrieval-oriented image transmission was designed to maximize the accuracy of the inference task, where the objective was defined as the cross-entropy between the ground truth and the predicted class (identity).
Also, in \cite{xie2021task}, a task-oriented semantic communication scheme was proposed, where the cross-entropy objective with multiuser multi-modal data fusion was considered.
In general, the task-relevant information $g(X)$ is  represented by deep semantic coding of black-box neural networks from the signals of various modalities, such as text, speech, image, and video streaming.

Alternatively, a principle of information bottleneck (IB) was introduced in \cite{tishby2000information} to characterize the relationships among $X$, $\Tilde{X}$, and $Y$ in terms of mutual information. 
The IB principle indicates that $\Tilde{X}$ should be sufficient for the inference task goal if the inequality (\ref{goaloriented}b) holds. 
In \cite{shao2021learning}, the IB principle was used for a communication system orienting image classification task. By combining IB with stochastic optimization, this method in \cite{shao2021learning} was then extended in \cite{pezone2022goal} to deal with the same task while simultaneously minimizing energy consumption and service latency.
For IoE applications with booming devices and data, distributed EL has been an appealing alternative technique to extend intelligent services from a centralized cloud data center to the proximity of edge nodes, allowing fast ML model training at edge while performing the task goal at the server \cite{Gunduz2020communicate}.
In a distributed IoE network, a lightweight DL-based semantic communication system was proposed in \cite{xie2020lite} by training semantic feature extraction and coding at edge nodes while updating at the server.
By applying the IB principle, a task-oriented communication design for cooperative inference by multiple edge devices was proposed in \cite{shao2021task}, where a group of edge nodes perform the inference task collaboratively with the assistance of an edge server.
It is concluded that EL methods are promising candidates for promoting goal-oriented semantic communications serving downstream tasks.

%% file: sections/section4-B.tex
\subsection{Wireless Sensing and Edge Caching}
Fast growing IoE devices impose heavy load to current wireless networks with limited spectral resource, requiring better energy utilization efficiency and low latency with high reliability.
Wireless sensing and edge caching are the promising solutions to cope with these challenges.
To develop these techniques for IoE applications with scalability and stability, EL plays an important role to help improve the performance of delay-sensitive sensing and spatially-variant caching tasks.

\subsubsection{Sensing}

Sensing is an essential function of IoE networks with vertical applications like Internet-of-Vehicles (IoV), UAVs, mobile robots, and smart city.
Particularly in an IoV network, collaboration among different types of networks is inevitable and the vehicles are usually equipped with sensors and radio transponders.
Such a multi-attribute network needs to process heterogenous sensing data.
Besides, it is difficult to leverage constrained resources such as energy, spectrum, and power to deal with high-mobility and severe noise in IoV.
To address these issues, EL releases a part of the computing and learning pressure to edge vehicle nodes, by which the computation load of the fusion center is greatly reduced. 
In addition, since the transmitted data in EL is preprocessed in edge vehicle nodes, communication overhead and processing delay of the entire IoV network can be effectively minimized\cite{zhou2021twolayer}.


Due to the characteristics of the edge nodes in sensing networks, there are numerous practical constraints in terms of bandwidth, computation, memory, and battery life.
To meet these constraints, one has to substantially reduce the requirement of communication accuracy and weaken the adaptability to dynamic IoE environments when applying EL with conventional DL methods like CNNs \cite{zhao2018deepthings}.
This problem was studied in \cite{imteaj2020fedar} for a distributed network of mobile robots with communications.
Also, in \cite{imteaj2020fedar}, FL was applied for monitoring device activities with local computing resource, and individual trust measures.
Especially during the training period, asynchronous FL was applied to accelerate the convergence when untrustworthy and ineffective devices were eliminated.

\subsubsection{Caching}
Besides the integration of sensing, edge data caching also plays a key role in future IoE networks. An exponential growth of data in smart cities is generated by massive smart devices, e.g., sensors, smartphones, autonomous vehicles, as illustrated in Fig. \ref{fig4first}. These explosive data could possibly saturate the traffic of wireless networks and prevent QoS from being satisfied \cite{khan2020edge}. For instance, as shown in Fig. \ref{fig4first}, the autonomous vehicles constantly communicate with both roadside sensors and adjacent autonomous vehicles to collect information about the environment \cite{chekired20195g}. Indeed, low latency design is a key challenge in this application. Edge data caching is an effective tool for alleviating high latency and heavy load on fronthaul networks. Repeated transmission of the same data can be avoided by caching the data at edge nodes. Along with the edge caching, a slew of notable challenges are the design of \emph{caching update policy} and \emph{caching transmission policy}.

In terms of \emph{caching update policy}, the freshness of cached data has a significant impact on the system caching update policy. For some practical scenarios, a new metric was proposed in \cite{kaul2012real}, that is, Age of Information (AoI). AoI is defined as the length of time since the last measurement of the data. Investigations on scheduling policies have been considered to minimize AoI by utilizing the queuing model and conventional optimization theory \cite{he2018optimal}. However, due to the lack of prior knowledge about network characteristics and data status, the queuing model is typically inapplicable in real-world network environments \cite{beytur2019age}. This has inspired the use of ML methods for edge caching, which are capable of recognizing dynamic situations of temporal variation. In \cite{hsu2017age}, an RL method was developed to find scheduling decisions with the goal of minimizing long-term AoI at a single edge node. The subsequent work in \cite{wu2020deep} dedicated to developing an energy-efficient caching update policy at a single edge node by using ML. Then in \cite{wu2021caching}, an intelligent caching policy was explored for multiple edge nodes under the coordination of a cloud. As a further step, broader performance metrics including AoI, energy consumption, fronthaul traffic, were examined, and a type of MARL, known as multi-agent discrete variant of soft actor-critic RL, was proposed to achieve caching update of multiple edge nodes \cite{wu2021caching}.

Another focus of edge caching is the \emph{caching transmission policy}. Due to the broadcast nature of wireless channels, interference is a critical challenge in cloud-edge caching IoE networks, which seriously affects the transmission rates of cached data. The study in \cite{deghel2015benefits} investigated joint optimization of caching and interference alignment under time-invariant channels. Since wireless channels are time-varying and cannot be modeled accurately, ML were further introduced to tackle this issue \cite{luong2019applications}. In \cite{he2017Deep}, a DRL algorithm was proposed to realize cache-enabled interference alignment. For a heterogeneous network, an extension of \cite{he2017Deep}, a caching, networking, computing integrated framework based on DRL was proposed to minimize the energy consumption \cite{he2017big}.

Edge caching not only improves transmission rates, but also reduces communication latency. Studies in \cite{zhou2016stochastic} and \cite{wei2018jointuser} considered the same objective of minimizing average communication delay. Yet, the former was to find the optimal caching data transmission policy by a conventional convex optimization tool, while the latter was to jointly optimize user scheduling and caching by a DRL algorithm. Further in \cite{zhong2020deep}, a multi-agent actor-critic algorithm, which is a type of multi-agent DRL (MADRL), was proposed to mitigate the transmission delay in decentralized edge caching.

%% file: sections/section4-C.tex
\subsection{Integrated Aerial and Satellite Computing Networks}

\subsubsection{Aerial Networks}
UAV is a promising technology for enabling B5G IoT which enhances the performance of edge networks by acting as an aerial BS. Typically, in an edge network, UAV with high mobility can assist mobile edge computing (MEC) in offloading computationally intensive tasks from IoT devices. The flexible deployed UAV can handle emergency communication in the cases of inevitable natural disasters or temporary malfunctioning. Recently, the integration of UAV into MEC systems based on a single-agent DRL has been studied from various aspects. For instance, an UAV path planning method based on the DRL was proposed in \cite{wan2019toward} to collect the distributed data from sensors for edge computing.

In future IoT edge networks, there can be multiple UAVs serving as multiple distributed edge agents. Then, MARL approaches have been exploited to solve the problems like distributed resource management for computation offloading at the edge network. For UAV-assisted edge computing networks, Sacco \emph{et al.} \cite{sacco2021sustainable} applied MARL to coordinately improve the system energy efficiency and accelerate task completion by distributedly offloading decision strategies. Also, in \cite{zhu2021learning}, Zhu \emph{et al.} proposed an MARL framework to learn the effect of environment on the offloading policy, where task allocation and bandwidth allocation are handled distributedly by two agents. In addition to MARL, DRL facilitates convergence by exploiting the power of DNNs for estimating the associated functions in conventional RL. For modeling cooperative computation offloading, an MADRL-based method was proposed in \cite{seid2021multi} to minimize the overall network computation cost. Meanwhile, in \cite{wang2020multi}, an MADRL-based trajectory control algorithm was developed to manage the trajectory design of each individual UAV in a decentralized manner.

These studies on MARL/MADRL focused on the design in a distributed manner without considering privacy protect of IoT devices. Recently, one of the latest researches considered the privacy issue by applying FL \cite{cheng2021intelligent}. Driven by the advantages brought by FL as described in the previous sections, a federated DRL (FDRL) framework was proposed to learn joint task offloading and energy allocation in an UAV-aided MEC system. More recently, in \cite{nie2021semi}, a semi-distributed multi-agent federated reinforcement learning (MAFRL) algorithm was devised to keep the data training locally and thus protect privacy of all IoT devices by the integration of FL and MADRL.

Besides being employed for computation offloading in MEC systems, UAV can also collaboratively perform AI tasks using their locally distributed data and computation capabilities. This provides a promising approach of meeting challenges of limited resources of edge devices as well as the ubiquitous coverage envisioned by B5G IoT. In \cite{zhang2021transfer}, distributed intelligence was delivered by UAVs to perceive environmental changes for edge service scheduling. In \cite{qu2021empowering}, a framework by integrating air-ground networks and FL was proposed to empower edge intelligence, where UAVs were deployed as aerial nodes to collaboratively train an effective learning model. Meanwhile, in \cite{lim2021uav}, UAVs were employed to provide intermediate model aggregation in FL models to improve the efficiency of both learning and communication. In addition, UAVs were also considered as edge servers for FL to boost edge intelligence in \cite{dong2021uavs} and were acting as wireless relays to facilitate the communications between vehicles and the FL server in \cite{ng2020joint}. Despite these research progress, fundamental performance limits of distributed edge learning with mobile UAV nodes is still a largely uncharted area.

\subsubsection{Satellite}
Thanks to the ability of providing seamless coverage for remote and depopulated areas, satellite communication forms a critical part of IoE in B5G networks \cite{shi2022intelligent}. In satellite-served IoE, computation tasks generated by terrestrial IoE devices are offloaded to satellites \cite{sthapit2021reinforcement}. These tasks can be processed by satellite-enabled local computation platforms or allocated to other compute nodes, such as space station, super computation satellite, and ground edge servers. However, limited by energy and computation capacity, satellite local compute servers cannot handle all tasks from IoE devices. If all of these tasks are offloaded to other compute nodes, high delay caused by queuing and transmission process may prevent these tasks from being processed in time. 
In addition, the network topology is dynamic and the channel fluctuation is fast due to high-speed movement of low earth orbit (LEO) satellites. All these impose significant challenges to the task offloading process.
Therefore, the computation offloading (CO) for satellite-served IoE requires complicated optimization involving energy consumption, computation delay, and computation capacity constraints.

ML-based edge computing is one of the vital enabling technologies for satellite-served IoE networks. It learns to offload computing tasks from ground terminals to multiple satellites and ground edge servers efficiently. In \cite{sthapit2021reinforcement}, the optimization problem of the CO policy design was established for satellite-served IoT network, which minimizes a weighted sum of delay, energy consumption, and safety risk factors. A DRL technology, called the deep deterministic policy gradient method, was leveraged in \cite{sthapit2021reinforcement} to solve this optimization problem. 

Considering that UAVs are closer to ground IoE devices than LEO satellites, they provide near-real-time computing service with less transmission power consumption. Satellite-UAV-served IoE networks are supplements to satellite-served IoE networks. The authors in \cite{mao2021opyimizing} adopted the DL technique of long short-term memory (LSTM) modules to predict the remaining energy of IoT devices. They utilized an AI-based method to design the task offloading policy according to  communication conditions and computation resources, aiming to maximize the number of completed tasks.
In \cite{cheng2019space}, the task offloading decision in a satellite-UAV-served IoT network was formulated as an MDP with network dynamics, and a DRL-based method was proposed to learn the optimal CO policy.

\begin{table*}[ht!]
\centering
\caption{Representative Works on EL}
\label{tableInSec4}
\begin{tabular}{|m{0.05cm}<{\centering}m{0.7cm}<{\centering}|m{1.5cm}<{\centering}|m{2.8cm}<{\centering}|m{3.8cm}<{\centering}|m{6.2cm}|}
\hline
\multicolumn{2}{|c|}{\textbf{AI Method}}                                    & \textbf{Reference} & \textbf{Application Scenario} & \textbf{Research Focus} & \multicolumn{1}{c|}{\textbf{Objective}} \\ \hline
\multicolumn{2}{|c|}{\multirow{3}{*}{DNN}}                         & \cite{shao2021learning},\cite{shao2021task} & Goal-oriented communication & Task inference   &  Enabling a task-oriented communication principle for edge device inference under the IB framework                              \\ \cline{3-6} 
\multicolumn{2}{|c|}{}                                             &     \cite{mao2021opyimizing}      &               Satellite-UAV-served IoE    &         Computation offloading        &                       Computation offloading policy to maximize the number of computing tasks          \\ \hline
\multicolumn{2}{|c|}{\multirow{6}{*}{{\begin{tabular}[c]{@{}c@{}}\\ \\ \\ \\ \\ FL \\ \\ \\ \\ \end{tabular}}}}                          &\cite{xie2020lite}         &   Distributed IoE            &  Semantics extraction              & Improving transmission efficiency with lightweight neural network  \\ \cline{3-6}
\multicolumn{2}{|c|}{}  & \cite{imteaj2020fedar} & Mobile robots & Learning process optimization & Dealing with unreliable and resource-constrained FL environment  \\ \cline{3-6} 
\multicolumn{2}{|c|}{}  &\cite{cheng2021intelligent,nie2021semi} & {\begin{tabular}[c]{@{}c@{}}UAV-aided MEC \end{tabular}}   & Resource allocation & Local data training with privacy protection                            \\ \cline{3-6} 
\multicolumn{2}{|c|}{}  &\cite{qu2021empowering,dong2021uavs,lim2021uav,ng2020joint} & {\begin{tabular}[c]{@{}c@{}}UAV-aided  MEC \end{tabular}}  & Edge intelligence & Collaborative model training                   \\ \cline{3-6} 
\multicolumn{2}{|c|}{}                                             
&\cite{sun2021dynamic,cao2021optimized,cao2020cooperative,zhang2021gradient} & {\begin{tabular}[c]{@{}c@{}}Over-the-air FL\end{tabular}}     &  Edge intelligence  &  Minimizing the average computation MSE                         \\ \hline
\multicolumn{1}{|c|}{\multirow{12}{*}{{\begin{tabular}[c]{@{}c@{}} \\ \\ \\ \\ \\ \\ \\ \\ \\ \\ RL \\ \end{tabular}}}} & \multirow{1}{*}{{\begin{tabular}[c]{@{}c@{}}  SARL \end{tabular}}} 
&\cite{hsu2017age} & {\begin{tabular}[c]{@{}c@{}}Edge sensing IoE \end{tabular}} & Caching updating policy  & Minimizing the average AoI and energy consumption    \\ \cline{2-6} 
\multicolumn{1}{|c|}{} & \multirow{2}{*}{{\begin{tabular}[c]{@{}c@{}} \\ MARL \\ \end{tabular}}} &\cite{sacco2021sustainable} & {\begin{tabular}[c]{@{}c@{}}UAV-aided MEC \end{tabular}} & Task offloading & Improving EE and accelerating task completion by distributed offloading          \\ \cline{3-6} 
\multicolumn{1}{|c|}{} &  &\cite{zhu2021learning}   & {\begin{tabular}[c]{@{}c@{}}UAV-aided MEC \end{tabular}} & Task offloading   & Minimizing the average mission response time for the inter-dependent tasks of dynamic UAVs \\  \cline{2-6} 
\multicolumn{1}{|c|}{}                     & \multirow{6}{*}{{\begin{tabular}[c]{@{}c@{}} \\ \\ \\ \\ \\ \\ DRL\\ \end{tabular}}} 
&    \cite{sthapit2021reinforcement,cheng2019space}        &               Satellite-and-UAV-served IoE        &         Computation offloading       &                         Finding the optimal computation offloading policy      \\ \cline{3-6}
\multicolumn{1}{|c|}{} &
& \cite{wu2020deep,wu2021caching} & {\begin{tabular}[c]{@{}c@{}}Edge caching IoE \end{tabular}} & Caching updating policy & Trading off between the average AoI and energy cost for multiple edge nodes \\\cline{3-6}
\multicolumn{1}{|c|}{} &
&\cite{he2017Deep,he2017big} & {\begin{tabular}[c]{@{}c@{}}Edge caching IoE \end{tabular}} & Caching transmission policy & Trading off between interference alignment, caching, and computing \\ \cline{3-6}
\multicolumn{1}{|c|}{} &
& \cite{wei2018jointuser,zhong2020deep} & {\begin{tabular}[c]{@{}c@{}}Edge caching IoE \end{tabular}} & Caching transmission policy &  Minimizing the average transmission delay \\ \cline{3-6}
\multicolumn{1}{|c|}{} &       &  \cite{wan2019toward}         & {\begin{tabular}[c]{@{}c@{}}UAV-aided MEC \end{tabular}}   & Big data processing              & Distributed path planning and resource management using single-agent DRL \\ \cline{3-6} 
\multicolumn{1}{|c|}{} &
& \cite{seid2021multi,wang2020multi} & {\begin{tabular}[c]{@{}c@{}}UAV-aided MEC \end{tabular}} & Resource allocation & Distributed multi-agent resource allocation in a multi-UAV enabled network  \\ \hline
\end{tabular}
\end{table*}

%% file: sections/section4-D.tex
\subsection{Over-the-air Computation}
Although numerous emerging applications at edge wireless networks have been developed to support universal connectivity and automatic processing, it is challenging to accomplish effective data aggregation for a huge number of edge devices. FL enables each device to upload model parameters obtained from local data training. To facilitate FL data collection from distributed devices, over-the-air FL (Air-FL) is envisioned to provide better performance with less bandwidth requirement \cite{zhu2018mimo}, \cite{li2019wirelessly}. Air-FL can be accomplished by exploiting functional decomposition and waveform superposition properties over multiple-access channels, i.e., the technique of AirComp. Early works on AirComp have concentrated on performance analysis and transceiver optimization regarding the average computation MSE. In particular, the comprehensive ergodic performance analysis and the average MSE minimization were considered in \cite{liu2020over}. In \cite{yang2021revisiting}, the authors modeled the statistics of interference and revealed a two-sided effect of interference on the overall training procedure in AirComp.

The advantages of Air-FL are indisputable, but they face two main practical limitations. First, the aggregation errors in Air-FL urgently need to be combated due to the feature of wireless fading channels, and second, the computation accuracy of Air-FL depends highly on the worst channel condition between the access point (AP) and the edge devices. In light of the first limitation, learning performance can be improved by selecting the optimal number of devices and optimizing the transmit power in Air-FL. For example, in \cite{yang2020federated}, efficient algorithms were proposed to cope with the nonconvex constraints for device scheduling and transmit beamforming. Further considering synchronization in Air-FL, dynamic scheduling of edge devices \cite{sun2021dynamic} and transmission power control \cite{cao2021optimized} were studied to mitigate the data aggregation errors. Meanwhile, a novel power control algorithm was presented to lighten the impact of inter-cell interference on performance in Air-FL \cite{cao2020cooperative}. In order to tackle the statistical characteristics of gradients which vary in each iteration, the transmit power control was obtained in closed form under fading channels \cite{zhang2021gradient}. Besides, since the training data of each user can also obey different distributions, a local SGD-based power control algorithm was proposed by introducing time-varying precoding \cite{sery2021over}.

On the other hand, the second limitation can be partially addressed through the advances of RIS to enhance the quality of the worst channel in Air-FL. Specifically, an alternative algorithm was presented to jointly optimize the passive beamforming at the RIS and active beamforming at the transmitter with low complexity, where the authors validate the superiority of integrating RIS into Air-FL \cite{fang2021over}. For the case of imperfect CSI, a robust design of RIS-assisted Air-FL was proposed in \cite{zhang2022worst} under a sum-power constraint. In addition, privacy in RIS-assisted Air-FL was studied based on differential privacy technique \cite{yang2022differentially}. As a consequence, one can tackle the above two limitations by considering device scheduling in RIS-assisted Air-FL. In \cite{Liu2021Reconfigurable}, the authors highlighted the effects of scheduling devices on training accuracy and demonstrated the necessity of optimizing RIS for achieving significant learning performance. To conclude, Air-FL has great potential to enhance the efficiency of data aggregation at each communication round for EL.

%% file: sections/section4-E.tex
\subsection{Holographic Beamforming}
The QoS provided by B5G IoE should satisfy the requirements of holographic communication, which requires uninterrupted high speed, low latency to constantly maintain virtual presence. As a result, EL provides a real-time processing paradigm with low latency which helps provide holographic contents to IoE devices on-demand. Recently, an intelligent augmented reality (Intelli-AR) preloading algorithm was proposed in \cite{han2022intelli} to improve the transmission efficiency in an edge network, where the edge servers proactively transmit holographic contents to IoE devices. The Intelli-AR solution is verified to improve the ratio of successful preloading by $11.52\%$ compared to the baseline in a practical dataset \cite{han2022intelli}. 

To fully unlock its potential, holographic communication needs to reach the complete control of the electromagnetic (EM) field generated and sensed by antennas. In such a case, EL is a supplement to cloud computing for the complex EM computing requirements in B5G IoE networks. A distributed DL algorithm can be carried out on edge nodes for computing tasks to reconfigure EM waves with reduced latency and energy consumption.


In summary, EL techniques have shown great potential in empowering these B5G wireless applications. For the convenience of comparison, representative works of these applications with EL are listed in Table~\ref{tableInSec4}.

%% file: sections/section5.tex
\section{Open Problems and Challenges}
In this section, we point out major open problems and challenges in implementing EL over B5G and IoE wireless networks.
\subsection{Open Problems in Information Theory}

In theory, the limits on information flow for communication networks were originally discovered by Claude Shannon in the seminal work \cite{shannon1948mathematical}. Most well-known information theoretical results thereafter successfully characterized the source information entropy and channel capacity of various communication networks for a conventional task of exactly recovering the source information at receiver(s) \cite{Cover2006ITbook}. For EL and inference, however, the task of a communication network can be the computation on a utility function, or an inference decision based on the distributed information, rather than recovering all source information explicitly. For instance, an FL server requests for only an averaging aggregation of the gradients calculated and sent by distributed ML nodes. Under these scenarios, the communication rate region established by existing distributed source coding theorems, e.g., the Slepian-Wolf theorem \cite{Slepian1973Coding} and the Wyner-Ziv coding \cite{Wyner1976Coding}, sometimes becomes much larger than the minimum communication rate needed for these computing tasks. Till today, the theoretical rate limits are available for a few types of computing tasks in very special use cases, e.g., the ``$\bm{\mu}$-sum'' computing task of two distributed Gaussian sources in a multiple access channel \cite{Gammal2011NetIT}. In particular, as we formulated the problem in \eqref{goaloriented}, it should be possible to design the optimal mapping function $g$ in a closed form, rather than by resorting to unexplained black-box neural networks, for a goal-oriented communication system with some specific task goals. Therefore, it is of both theoretical and practical importance to find the rate-tuple limit of distributed source coding for general computing tasks in B5G networks with heterogeneous data and arbitrary topology.
    
On the other hand, current design of separate source-channel coding in most communication systems is guaranteed optimal in theory under the assumptions of using long codewords and aiming at exact information recovery. In B5G networks, it is natural to expect growing demands on short-packet data communication by the applications of distributed FL and sensing data collection \cite{Latief2021EdgeAI}. For these short-packet transmissions, source-channel separation can be far from optimal to fulfil the edge computing and learning tasks, even if the distributed sources are independent \cite{Gammal2011NetIT}. To improve the performance, state-of-the-art goal-oriented communication designs have advocated joint source-channel coding techniques using DL. However, these data-driven DL techniques have to be trained case-by-case for vast applications. Also, their performance gaps to the optimum are still unknown in general. In order to promote the widespread goal-oriented semantic communication design in practice, we should envision significant benefits of completing theoretical studies on joint source-channel coding for EL and computing. In addition, related theoretical development should provide an extra potential dimension of joint source-channel coding to strengthen stream data caching, privacy, and communication security in EL networks.

\subsection{Architecture and Technique Challenges}
The architecture of EL mainly faces two kinds of challenges, i.e., from the communication system and learning procedure. 
As for the communication system, B5G networks tend to be heterogeneous with multiple tiers of BSs. 
Besides, due to the mobility of edge devices, they can leave or join the communication system. 
Thus, one challenge is that EL should be adaptive to the time varying heterogeneous properties of network as well as the mobility of edge devices. 
As for learning procedure, the EL performance can only be well guaranteed when the dataset is  uniformly distributed among edge devices.
However, edge devices usually have non-i.i.d. dataset, which can lead to poor learning performance of EL.  
As a result, the other challenge is that model aggregation of EL should cope with non-i.i.d. dataset. 

It is also of great interest  to investigate the joint communication and computation resource allocation, such as communication bandwidth
for improving convergence of FL. FL relies on mobile wireless communications to collaboratively
learn ML models. Although the computing resource of mobile phones is
becoming more powerful, the bandwidth of wireless communication has not
increased much. Therefore, the bottleneck shifts from computing to communicating. The limited communication bandwidth may cause a longer communication delay,
which definitely results in longer convergence time in FL.
\subsection{Research Opportunities with New Applications}
In B5G communication networks, there are new emerging applications such as blockchain techniques \cite{li2020blockchain}, quantum computing \cite{chehimi2021quantum}, and metaverse.
Since the central learning model can suffer from servers' constant attack  and there can exist malicious clients, 
the security is an important issue of EL. 
Combining the committee consensus mechanism of blockchain technique, EL framework can effectively reduce consensus  computation and malicious attacks.
Due to the explosive growth of data edge devices,  quantum computing can be effectively utilized to solve large complex EL problems through performing classic ML tasks on quantum data.

With the rapid development of wireless networks and AI, emerging applications continue to appear and metaverse is a future perspective of wireless communication systems to realize a virtual digital universe. Emerging metaverse applications have put forward higher demands on 6G networks for end-to-end information processing capabilities. In order to meet these higher performance demands, 6G will be an end-to-end information processing and service network, and its core functions will expand from information transmission to information collection, information computing and application, and providing stronger sensing, communication, and computing capabilities.
It is also expected that advancements in joint sensing, communication, and computing would help form a platform for implementing EL in B5G networks.

%% file: sections/section6-conclusions.tex
\section{Conclusion}

In this paper, we  presented a comprehensive overview on distributed EL techniques.
We introduced the interplay between EL and communication optimization design. 
In particular, we provided dual-functional performance metrics for both learning and communication.
We also pointed out the communication optimization design for EL and learning techniques for communication optimization from the pointview of signal processing. 
Moreover, we provided the detailed B5G applications, open problems, and challenges of EL framework.  
The in-depth study on the signal processing techniques for the EL over wireless communications provides guidelines for the native integration of ML and edge networks.